\documentclass{aa}  

\usepackage{graphicx}	
\usepackage{amsmath}	
\usepackage{amssymb}	
\usepackage{multirow}
\usepackage{lscape}
\usepackage{makecell}
\usepackage{subfigure}
\usepackage{float}
\usepackage[colorlinks=true, allcolors=blue]{hyperref}

\usepackage{txfonts}
\begin{document}

   \title{Transient QPOs of Fermi-LAT blazars under the Curved Jet Model}

   \author{P. Pe\~nil\inst{1}
          \and
          J. Otero-Santos\inst{2}
          \and
          A. Banerjee\inst{1}
          \and
          S. Buson\inst{3,4}
          \and
          A. Rico\inst{1}
          \and
          M. Ajello\inst{1}
          \and 
          S. Adhikari\inst{1}
          }

   \institute{
            Department of Physics and Astronomy, Clemson University, Kinard Lab of Physics, Clemson, SC 29634-0978, USA\\
            \email{ppenil@clemson.edu}
         \and
             Istituto Nazionale di Fisica Nucleare, Sezione di Padova, 35131 Padova, Italy\\
             \email{jorge.otero@pd.infn.it}
        \and 
             Julius-Maximilians-Universit\"at W\"urzburg, Fakultät f\"ur Physik und Astronomie, Emil-Fischer-Str. 31, D-97074 W\"urzburg, Germany\\
        \and
            Deutsches Elektronen-Synchrotron DESY, Platanenallee 6, 15738 Zeuthen, Germany
             }

   \date{Received XX, XX; accepted July 3, 2025}

 
  \abstract
   {This study explores transient quasi-periodic oscillations (QPOs) in the $\gamma$-ray emission of two blazars, PMN J0531$-$4827 and PKS 1502+106, using over a decade of Fermi Large Area Telescope observations.}
   {The analysis focuses on identifying QPO signatures in their long-term light curves and interpreting the variability through a curved jet model, which predicts multiplicative oscillations with exponentially decaying amplitudes.}
   {We develop an analysis methodology to characterize the QPO and the specific properties of the amplitude of such QPOs.}
   {The findings offer insights into the dynamic processes driving relativistic jet evolution and their potential connections to underlying mechanisms, such as binary systems or other phenomena influencing the observed characteristics of these blazars.}
   {}


   \maketitle
%

\section{Introduction}

Blazars, a class of active galactic nuclei (AGN) characterized by powerful relativistic jets oriented close to our line of sight \citep[e.g.,][]{wiita_lecture}, are known for being the main population of extragalactic $\gamma$-ray emitters \citep[e.g.,][]{ghisellini1998}. Their emission exhibits complex variability across a wide range of timescales, from minutes to years \citep[e.g.,][]{fan2005, penil_mwl, raiteri_2024}. Among the diverse variability patterns, quasi-periodic oscillations (QPOs) have attracted increasing interest due to their potential to reveal fundamental aspects of jet physics and black hole accretion processes \citep[e.g.,][]{otero_mwl, sagar_pg1553}. QPOs in the $\gamma$-ray emission of AGN are particularly intriguing, as these quasi-periodicities may be linked, among other scenarios, to oscillatory mechanisms within the relativistic jet structure \citep[e.g.,][]{camenzind_jet, pg1553_precession} or the dynamics of the central supermassive black hole (SMBH) system \citep[e.g.,][]{sandrinelli_redfit, sobacchi_binary}.

Transient QPOs are a type of QPO characterized by their brief duration and episodic appearance within the light curves (LCs) of blazars \citep[][]{sarkar_transient_qpo, roy_transient_qpo}. The detection of transient QPOs in blazars holds significance in advancing our understanding of AGN emission mechanisms, as these events often signal underlying periodicity associated with specific physical processes. However, such transient QPOs can stem from turbulent processes within the jet or the broader AGN environment, which are inherently stochastic \citep[e.g.,][]{ruan_stocastic, covino_negation}, resulting from red-noise processes \citep[][]{vaughan_criticism}. These transient modulations in $\gamma$ rays have been reported in several blazars \citep[][]{gong_transient_qpos, chen_transient_qpo}. 

Different interpretations have been proposed to explain transient QPOs in the $\gamma$-ray emission of blazars using theoretical models of relativistic jets. One explanation involves a precessing jet \citep[][]{camenzind_jet}, where the precession could result from different mechanisms, such as Lense-Thirring precession of the accretion disk \citep[][]{fragile_meier_lense, franchini_lense} or the orbital motion of a binary SMBH \citep[][]{valtonen_oj287, qian_precesion_binary}. These dynamic processes could produce periods of $\gtrsim$ one year \citep[][]{rieger_2004}. Another hypothesis suggests that transient QPOs arise from turbulence mechanisms within the jet, particularly in regions influenced by moving or stationary shocks \citep[e.g.,][]{marscher_1992}. Given the stochastic nature of turbulence, such oscillations are unlikely to persist over multiple cycles.

In this work, we explore QPOs in the $\gamma$-ray LCs of two blazars, based on data collected by the Fermi Large Area Telescope \citep[LAT,][]{fermi_lat}. These QPOs are short-lived, typically occurring over a limited number of cycles and spanning several years. We interpret such transient oscillations based on a curved jet model to explore whether the transient periodicities observed can be attributed to the changing orientation of the jet \citep[e.g.,][]{sarkar_curved_jet,anuvab_curve_jet}. In the curved jet model, the trajectory of the jet follows a curved path. This jet structure has been observed in some AGN by using radio observations \citep[e.g.,][]{lister_vlbi_bent_jet}.

The structure of this paper is organized as follows. In Sect.~\ref{sec:sample}, we introduce the sources analyzed in this study and discuss the Fermi-LAT data utilized. Sect.~\ref{sec:methodology} details the methodology followed to characterize these QPOs. In Sect.~\ref{sec:model}, we describe the theoretical model applied to interpret the QPOs observed in our blazar sample. In Sect.~\ref{sec:flux_analysis}, we present and discuss the results obtained for each blazar. Finally, Sect.~\ref{sec:summary} summarizes our main findings and conclusions.

\section{Sample}\label{sec:sample}
The sample selection criteria relied on the 1,492 variable jetted AGN contained in the 4FGL-DR2 catalog \citep[][]{4fgl_dr2}, constructed with the first 12 years of data from the Fermi-LAT telescope. This sample was analyzed for searching long-term trends (an increase or decrease in flux over time) in the $\gamma$-ray LCs \citep{penil_2025}. Out of the 1,492 objects, 494 have also been analyzed by \cite{alba_ssa} (selected based on brightness and variability criteria) using a novel tool for time series analysis described in Sect.~\ref{sec:methodology}.

\begin{figure*}
	\centering
        \includegraphics[scale=0.4]{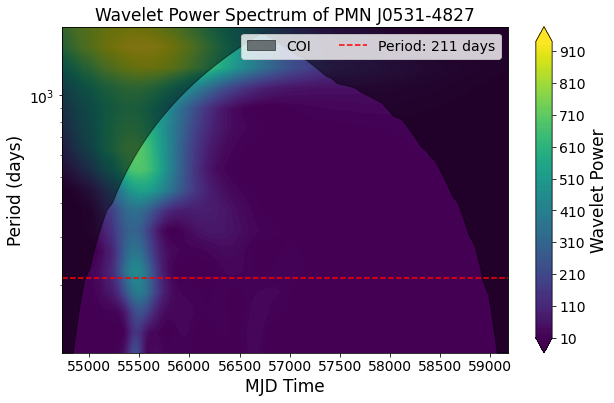}
        \includegraphics[scale=0.4]{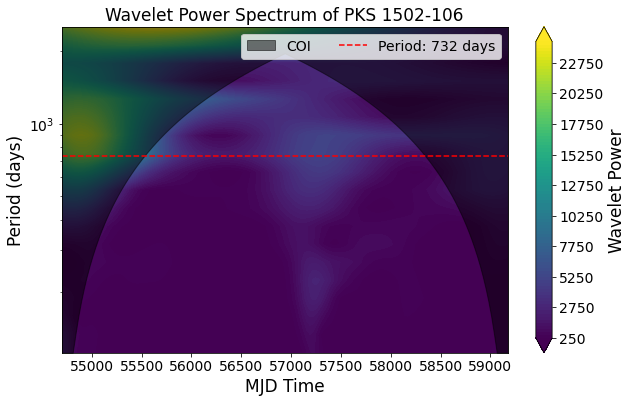}        
	\caption{Wavelet spectra of the complete $\gamma$-ray LCs of PMN J0531$-$4827 (Left) and PKS 1502+106 (Right). The colormap represents the signal's power spectrum, with color intensity indicating the strength of QPO components at different timescales. In this case, yellow colors represent higher power, suggesting stronger QPO signals, while purple corresponds to weaker or insignificant variations. This visualization identifies patterns and scales of variability and highlights any QPO presence. The shaded area, known as the cone of influence (COI), marks regions where potential modulations are either too close to the sampling interval or the signal's total duration, reducing reliability. The red horizontal line indicates evidence of transient QPOs. For PMN J0531$-$4827, a potential transient QPO appears around MJD 55500, with a similar occurrence noted for PKS 1502+106.} \label{fig:qpos_cwt}
\end{figure*}

We performed a systematic periodicity analysis using the pipeline described in \citet{penil_2020}, which resulted in a total of 998 sources that did not exhibit statistically significant long-term quasi-periodic emission, defined as having a period with a significance of $\leq 1.5\sigma$. Then, a cross-identification was performed between this subsample and the 337 objects that presented signatures of long-term increasing and/or decreasing trends studied in \citep{penil_2025}. This cross-match between both subsamples results in a total of 93 identified sources potentially showing both characteristics, that is, potential transient QPO signatures and trends in their $\gamma$-ray emission.

\begin{figure*}
	\centering
        \includegraphics[scale=0.21]{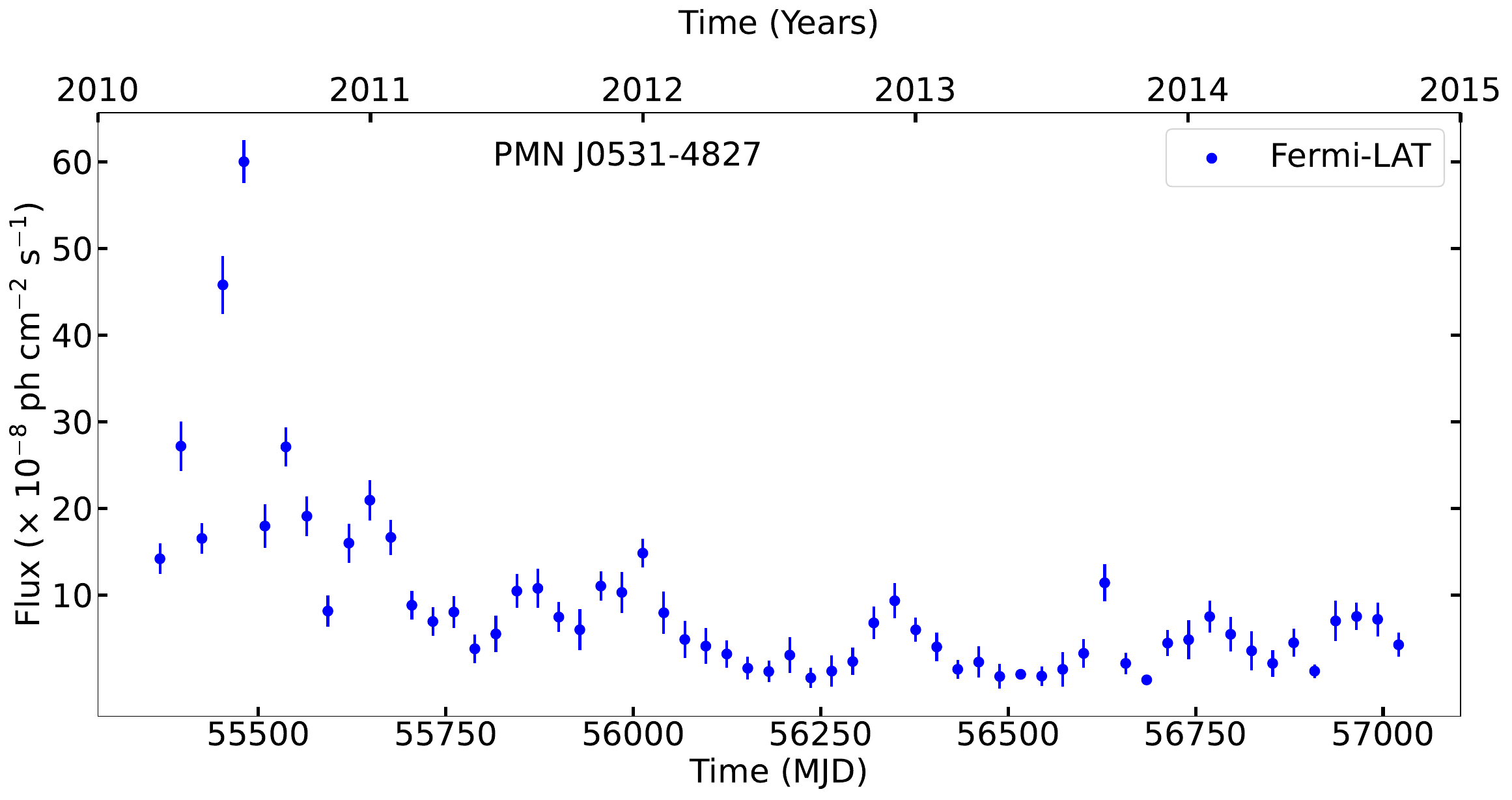}
        \includegraphics[scale=0.21]{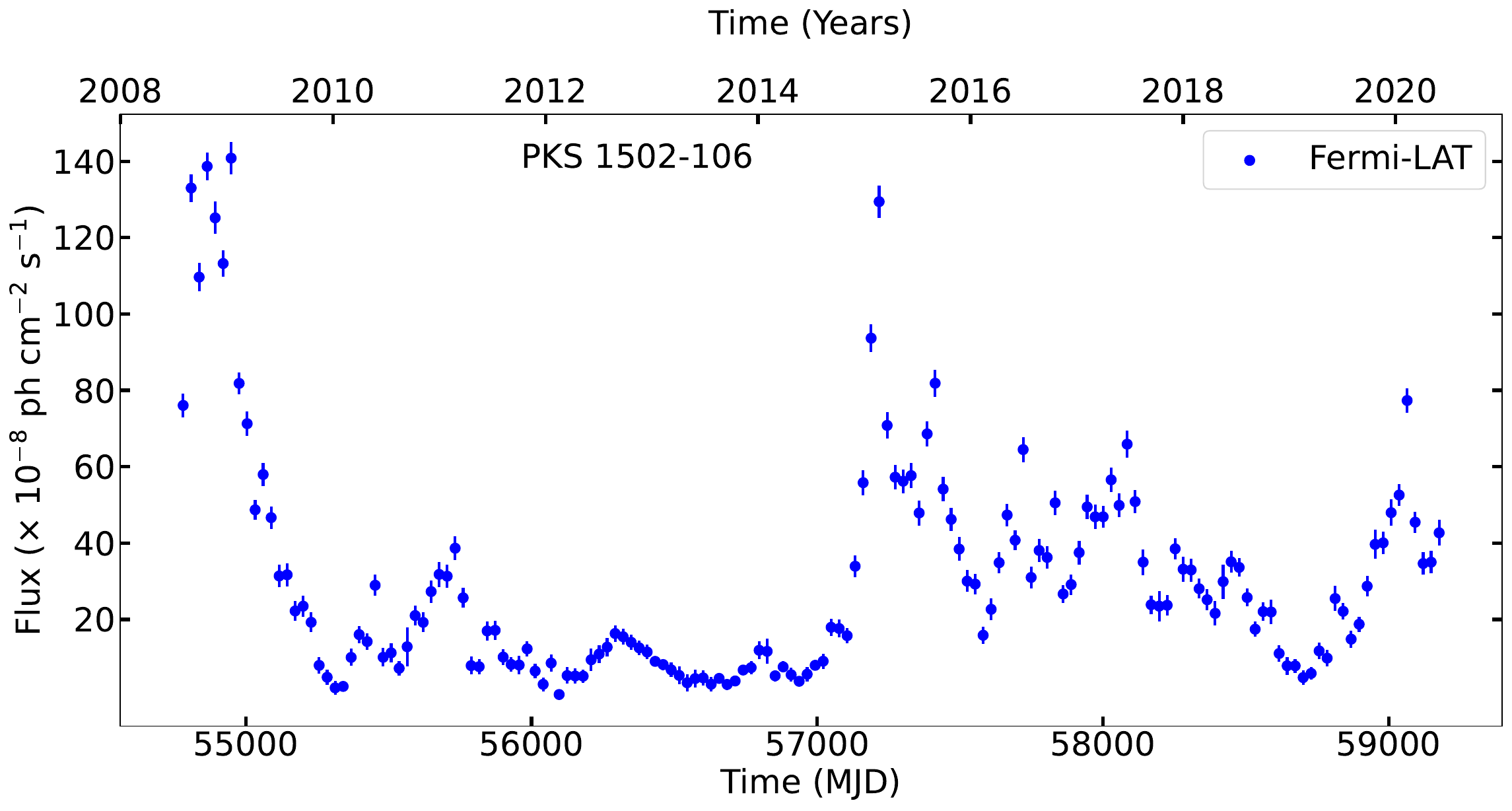}
	\caption{LCs where the CWT analysis denotes the presence of a potential transient QPO. Left: the segment of the LC of PMN J0531$-$4827 with a potential presence of QPO. Right: $\gamma$-ray LC of PKS 1502+106. These LCs are further analyzed in detail in Sect.~\ref{sec:flux_analysis}, where we investigate the temporal behavior and perform theoretical modeling in Sect.~\ref{sec:model}}. \label{fig:qpos_lcs}
\end{figure*}

These 93 sources are further studied with a detailed periodicity analysis using the Continuous Wavelet Transform \citep[CWT, see][]{wavelet_torrence}. Time series tools based on wavelet methods, such as the CWT, allow us to simultaneously transform the data into the time and frequency domains, identifying possible periodic variations and determining their persistence over time. Therefore, they are ideal for evaluating the presence of transient QPOs and finding the specific LC segments where they may occur (see Fig. \ref{fig:qpos_cwt}). Finally, among the studied blazars, we select those showing properties similar to those studied in \citet[][]{sarkar_curved_jet, anuvab_curve_jet}; specifically, sources that show potential QPOs with decreasing multiplicative amplitudes. Based on these criteria, we identified two blazars with these characteristics showing promising signatures of QPOs in their LCs (see Fig. \ref{fig:qpos_lcs}): 

\begin{itemize}
    \item PMN J0531$-$4827 (J0532.0-4827): BL Lacertae (BL Lac), $z=0.8116$ \citep{titov2017}.
    \item PKS 1502+106 (J1504.4+1029): flat-spectrum radio quasar (FSRQ), $z=1.839$ \citep{akiyama2003}.
\end{itemize}

PKS 1502+106 was proposed as a neutrino candidate by \citet[][]{britzen_precission_2016}, who identified a periodicity of approximately 3 years in $\gamma$-ray emission within a time frame similar to our observations. However, their study did not include a statistical significance analysis, highlighting the need for future investigation. They considered the possibility of a binary SMBH system as the driving mechanism behind the observables. 

The Fermi-LAT data analysis was performed following the procedure detailed in \citet{penil_2022_periodicity, alba_ssa}. In particular, we have used a binned likelihood analysis considering all data $\geqslant$0.1 GeV. The data were binned in 28-day time bins, allowing us to study the long-term behavior of the emission of the selected sources. Their LCs are shown in Fig. \ref{fig:qpos_lcs} and Fig. \ref{fig:complete_lc_pmn}.

\section{Methodology} \label{sec:methodology}
This section details the methodology employed to analyze the LCs and identify their key properties, providing the foundation for the flux analysis of each blazar. 

\subsection {Methods for the Search of QPOs}
After the first screening performed during the source sample selection (see Sect.~\ref{sec:sample}), we base QPO evaluation on two different techniques, the Singular Spectrum Analysis \citep[SSA,][]{ssa_greco, SSA_algorithm} and the Generalized Lomb-Scargle Periodogram \citep[GLSP,][]{lomb_gen}.

\begin{figure*}
	\centering
        \includegraphics[scale=0.21]{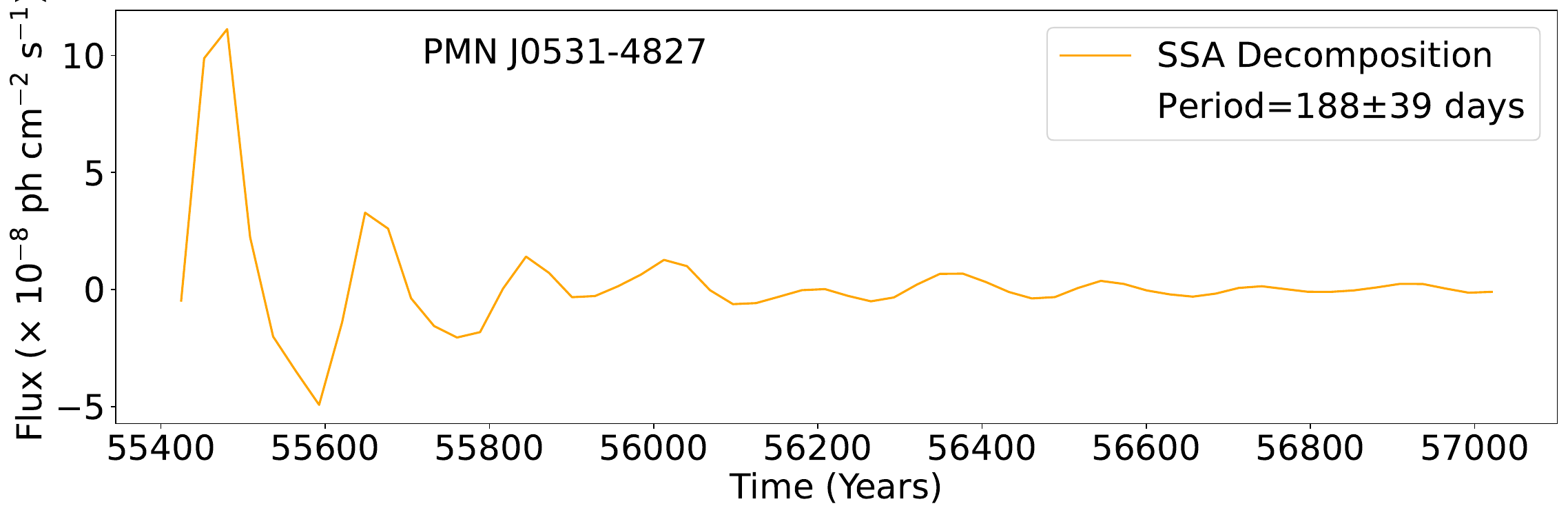}
        \par\vspace{0.5cm}
        \includegraphics[scale=0.21]{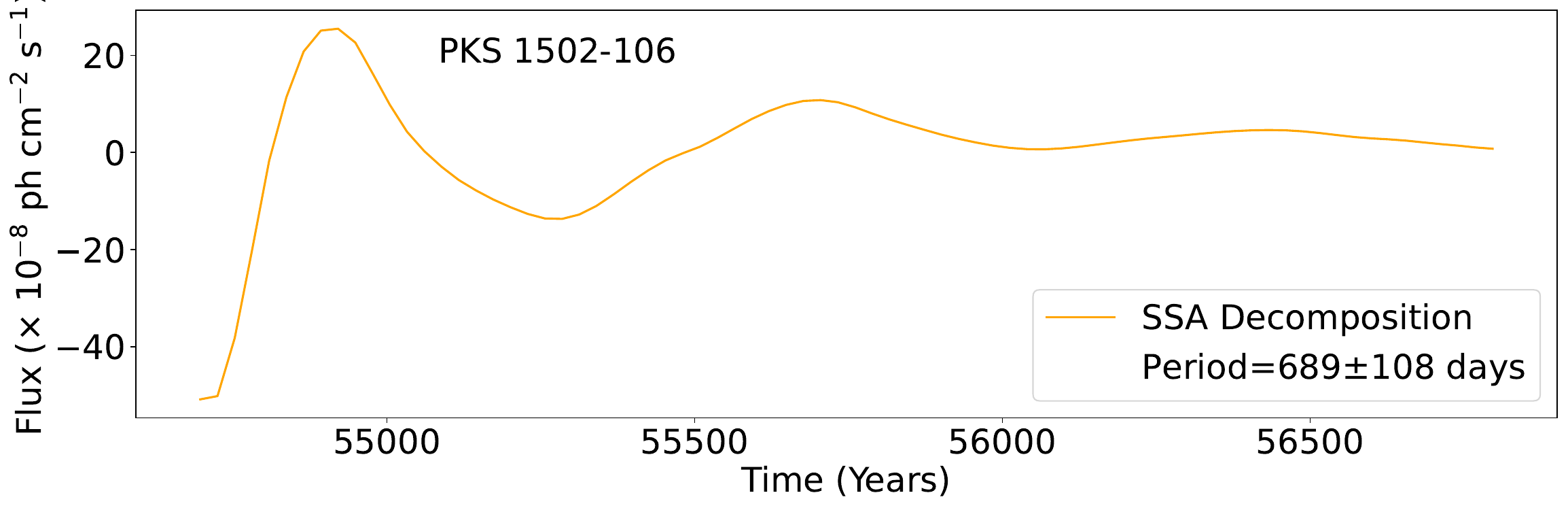}
        \includegraphics[scale=0.21]{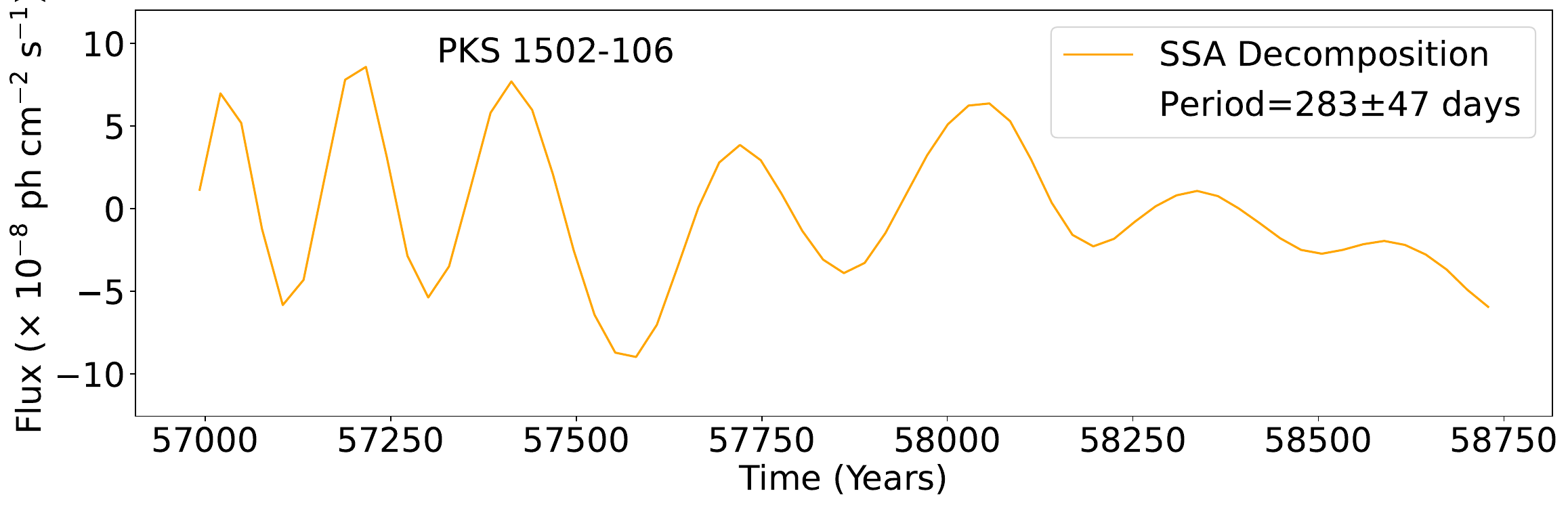}
	\caption{SSA decomposition of LCs of the blazars, revealing the oscillatory component in the signal. Top: PMN J0531$-$4827, and Bottom: PKS 1502+106, showing the decomposition os the two segments analyzed of the LC. The oscillatory patterns extracted from the LCs highlight a periodic component, with the inferred periods and uncertainties. The SSA decomposing follows the steps detailed in \citet{alba_ssa}.} \label{fig:qpos_ssa}
\end{figure*}

SSA is a non-parametric approach that breaks down a time series into its main components, being able to identify trends, oscillatory patterns, and noise (see Fig. \ref{fig:qpos_ssa}). This method is particularly effective for identifying quasi-periodic signals embedded within noisy datasets \citep[][]{alba_ssa}. Unlike many other techniques, SSA does not rely on predefined assumptions about periodicity and does not require the signal to be sinusoidal, making it a versatile and reliable tool for detecting QPOs with diverse frequencies and amplitudes.  

The second method, GLSP, built on the traditional Lomb-Scargle Periodogram \citep[LSP,][]{Lomb_1976, Scargle_1982}, is a commonly applied tool in periodicity studies dealing with unevenly-sampled, red-noise dominated datasets. The GLSP enhances the original method by accommodating non-sinusoidal periodic signals and integrating weights based on the uncertainties of individual data points, improving its sensitivity and accuracy.
This method is applied as a second step after the SSA on the isolated oscillatory component, when any, determining its period and statistical significance over the dominant red noise, following the methodology outlined by \citet{alba_ssa}.

To obtain an approximate estimate of the uncertainty in the derived periods from GLSP and SSA, we adopt the half width at half maximum of the corresponding peaks \citep[e.g.,][]{otero_mwl}.

\subsection {Test Statistics of the QPOs}
Identifying QPOs in time-series data is often complicated by noise, which can obscure or hinder genuine periodic signals. In blazar LCs, the dominant red noise contribution frequently leads to false detections \citep[][]{vaughan_criticism}, as it often mimics the expected behavior of a genuine periodic signal during a few cycles. This, combined with the Poisson white-noise-like component arising from the random photon detection process that particularly affects $\gamma$-ray LCs, is the main factor affecting the detectability of real (quasi-)periodic signals in AGN time series. 

\begin{figure}
	\centering
        \includegraphics[scale=0.22]{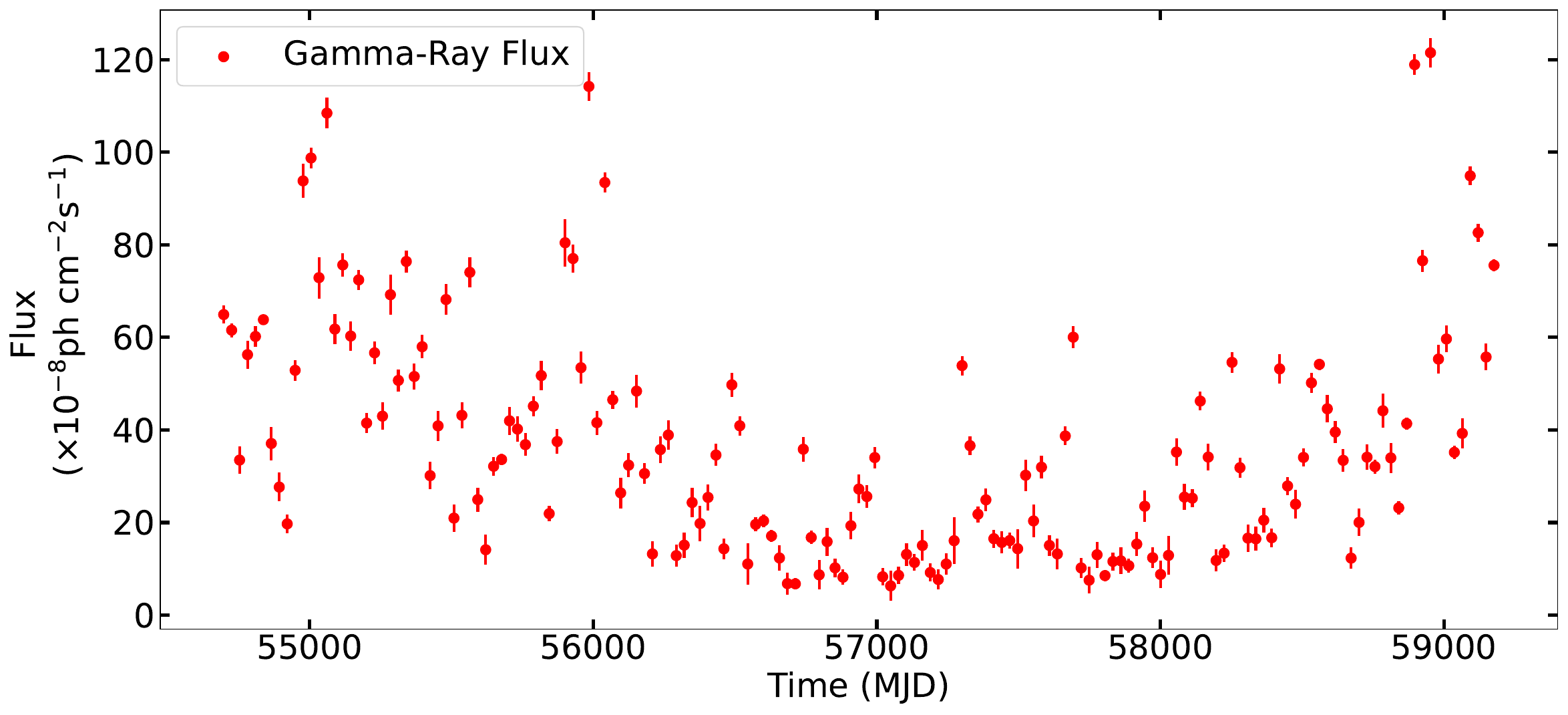}       
	\caption{Example of a synthetic LC used to estimate the significance associated with the periods obtained from SSA and GLSP analyses.}. \label{fig:synthetic_lcs}
\end{figure}

To assess the statistical significance of the QPOs discussed in this work, we adopt the procedure described in \citet{penil_2022_periodicity}. Following the methodology from \citet{emma_lc}, we generate 150,000 synthetic stochastic LCs that reproduce both the power spectral density (PSD) and the probability distribution function of the observed data (see the example shown in Fig. \ref{fig:synthetic_lcs}). These simulations provide a large ensemble of surrogate LCs that serve as a noise-only baseline, allowing us to evaluate how often the peaks in the observed QPOs could arise purely from stochastic variability. The resulting distributions from these synthetic datasets are then used to estimate the confidence levels (quoted as $\sigma$ values in the next section) associated with the observed QPOs.

The PSD is modeled as a power-law function, $A*f^{-\beta}+C$, where $A$ is the normalization, $\beta$ is the spectral index, $f$ is the frequency, and $C$ accounts for the Poisson noise component. The best fit is obtained through a Maximum Likelihood Estimation method and refined through Markov Chain Monte Carlo (MCMC) analysis\footnote{We utilize the \textsc{Python} package \texttt{emcee}.}. This approach allows for a robust characterization of the noise properties of our data, improving the reliability of QPO detections. 

\subsection {Results}
The results derived for the potential QPOs of the two sources analyzed here with the methodology described above are summarized in Table~\ref{tab:fitting_results}. For PMN J0531$-$4827, the observed QPO period is 188$\pm$39~days (2.7$\sigma$) using SSA and 213$\pm$46~days (1.1$\sigma$) with GLSP, showing consistent results within uncertainties. It is important to highlight that both methods show notable differences in derived significance, suggesting that SSA effectively reduces the contribution of noise components by isolating the genuine QPO, enhancing its detectability with respect to the GLSP.

For PKS 1502+106, the estimated period is 689$\pm$108 days (2.4$\sigma$) for SSA and 661$\pm$72 days (2.1$\sigma$) for GLSP. Again, both methods show a very close agreement between the reported values, in this case with a comparable significance level, suggesting a robust periodicity detection. The similarity in results across both methods indicates that the periodic signal is likely intrinsic rather than an artifact of the analysis technique. However, the moderate significance levels imply that the periodicity may still be influenced by the decay in the amplitudes, as visible in Fig.~\ref{fig:qpos_lcs}, or the reduced number of oscillations of the QPOs.   

\section{Theoretical Model} \label{sec:model}
\begin{figure}
	\centering
        \includegraphics[scale=0.41]{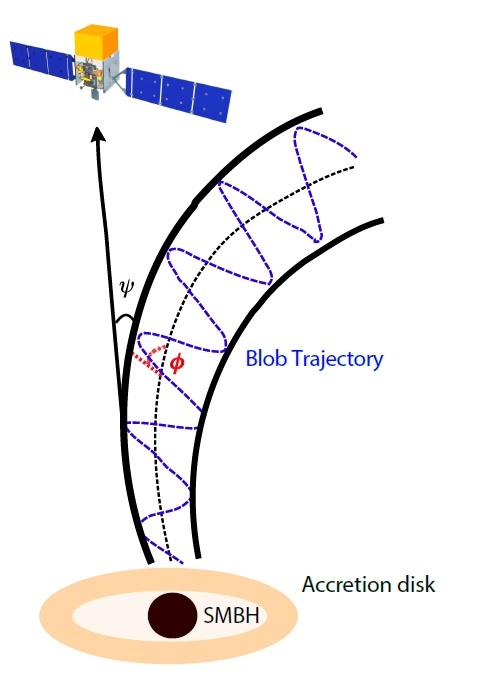}
	\caption{Sketch of the curved jet model based on \citet{sarkar_curved_jet}. The angle $\phi$ represents the inclination between the jet axis and the direction of the plasma blob's motion, while $\psi$ denotes the angle between the jet axis and the observer's line of sight. The dotted blue line traces the theoretical trajectory of the plasma blob within the jet, depicting its motion as influenced by the jet's curvature.} \label{fig:curved_jet_model}
\end{figure}

Several scenarios and hypotheses have been proposed in recent years to interpret QPOs in the emission of AGN and blazars. One of them, which we adopt in this work, is based on plasma blobs following helical paths within a curved jet \citep[e.g.,][]{sarkar_curved_jet, anuvab_curve_jet}. In this scenario, the blobs are ejected from the central region, often as a result of instabilities in the accretion disk or magnetic reconnection events near the SMBH \citep[][]{mohan_mangalam_blobs}. These blobs travel outwards along the curved jet, maintaining a helical trajectory due to the combined effects of the geometry of the jet and the magnetic field structure. Their motion can then produce quasi-periodic enhancements of the emission whenever they traverse regions of the jet that align with the observer’s line of sight and the opposite effect when the alignment is lower (Fig. \ref{fig:curved_jet_model}). In these instances, relativistic beaming modifies the observed intensity. However, as the blob moves within the curved jet, the oscillatory signal diminishes over time, ultimately fading \citep[e.g.,][]{sarkar_curved_jet, anuvab_curve_jet}. 

Several interpretations have been proposed for the origin of this scenario. \citet{rieger_2004} explored how differential Doppler boosting, caused by the helical motion of emitting components within the jet, can lead to observable periodicity. This motion is nonballistic, meaning the radiating plasma elements follow curved trajectories rather than moving in straight lines. \citet{rieger_2004} identified three possible physical mechanisms for helical jet motion in blazars. The first is nonballistic motion driven by the orbital motion in an SMBH binary, and thus, the QPO is linked to the orbiting motion of the binary. The second involves either ballistic or nonballistic helical jet paths induced by jet precession, which may result from gravitational torques exerted by a secondary black hole or from accretion disk warping. The third mechanism attributes nonballistic helical motion to intrinsic jet instabilities. The first two scenarios are particularly viable for explaining $P_{obs} \gtrsim$1 year. 

Alternatively, distortions in the jet structure could result from the Lense-Thirring effect, where frame-dragging caused by a rotating SMBH induces precession in a tilted accretion disk. This precession propagates down the jet, potentially causing it to bend \citep[e.g.,][]{caproni_spin_jet_precession}. A warped accretion disk, driven by gravitational torques or differential radiation pressure, can also induce global precession down to the structure \citep[e.g.,][]{liska_precession}.

In another scenario, magnetic fields interacting with the plasma in the accretion disk exert torques that may align or misalign sections of the disk, depending on the system’s initial configuration \citep[e.g.,][]{mcKinney_precession}. When the disk is misaligned with the SMBH’s spin axis, magnetic torques align the inner disk with the black hole’s equatorial plane, while the misaligned outer disk experiences differential torques. This leads to the precession of the disk plane, which propagates outward and causes changes in the jet’s orientation.

The bending of the jet can also be attributed to the magnetic pressure exerted by an oblique field, which acts asymmetrically on the jet and causes it to deviate from its initial trajectory \citep[][]{koide_jet_bent}. Alternatively, velocity shear between the jet and the surrounding medium can induce Kelvin-Helmholtz instabilities, leading to the formation of helical distortions and potential jet bending \citep[][]{Hardee_jet_bent}. Finally, compact objects such as stars, stellar remnants, or dense molecular clouds passing near or through an AGN jet can significantly affect its properties. Their strong gravitational influence may deflect the jet, cause fragmentation, or alter its collimation \citep[e.g.,][]{torres_jet_obstacles}. In this context, \citet{raiteri_2024} proposed a jet model incorporating an inhomogeneous, curved, and twisting structure, where a wiggling, filamentary jet, interpreted as a rotating double-helix, reproduces the multi-wavelength variability of BL Lacertae.

\subsection{Model expressions} \label{sec:equations_model}
Due to the close alignment of the jet and the line of sight, the emission of blazars is highly enhanced through relativistic Doppler boosting. The relativistic Doppler factor $\delta$, quantifying this enhancement, can be expressed as

\begin{equation}
\label{eq:delta}
\delta=\frac{1}{\Gamma[1-\beta\cos(\theta)]},
\end{equation}
where $\Gamma$ is the Lorentz factor, $\beta$ corresponds to the velocity of the jet in units of the light speed, and $\theta$ represents the viewing angle. As introduced above, the helical motion of the blobs along the jet results in changes in the viewing angle. According to Eq.~(\ref{eq:delta}), this results in variations in the radiation boosting and, therefore, the observed flux. The helical nature of this trajectory gives place to recurrent changes of the viewing angle $\theta_{obs} \left( t \right)$, inducing the observed quasi-periodic behaviors. This, combined with the curved morphology of the jet, can lead to the transient, fading QPOs observed in these sources. Under this scenario, we can express the viewing angle $\theta_{obs} \left( t \right)$ as a function of the angle between the jet and the direction of the plasma blob --- known as pitch angle $\phi$ --- and the angle between the jet and the line of sight $\psi$, as \citep[see e.g.][]{sarkar_curved_jet} 

\begin{equation}
\label{eq:cos_theta_time}
\cos\theta_{obs}(t) =\cos(\phi)\cos(\psi)+\sin(\phi)\sin(\psi)\cos(2\pi t/P_{obs}).
\end{equation}
Moreover, we can express the emitted flux as

\begin{equation}
    \label{eq:flux}
    {F_{\nu} \propto F_{\nu^{'}}^{'}}{\delta^{-(n+\alpha)}}.
\end{equation}
$F_{\nu}$ and $F_{\nu^{'}}^{'}$ correspond to the observed and rest-frame flux, respectively, $n$ represents the Doppler boosting factor and $\alpha$ is intrinsic spectral index. Taking into account  Eqs. (\ref{eq:delta}) and  (\ref{eq:cos_theta_time}), we can express the sources' flux as

\begin{equation}
    \label{eq:flux_expression}
    \begin{split}
        F_{\nu} \propto  \frac{F_{\nu^{'}}^{'}}{\Gamma^{(n+\alpha)} (1 + \sin(\phi) \sin(\psi))^{(n+\alpha)}} \times \\
        \left[ 1 - \frac{\beta \cos(\phi) \cos(\psi)}{1 + \sin(\phi) \sin(\psi)} 
        \cos\left(\frac{2\pi t}{P_{obs}}\right) \right]^{-(n+\alpha)}.
    \end{split}
\end{equation}

This emission model is therefore characterized first by the helical motion of the blob along the jet causing the observed recurrent variations and second by the change of the viewing angle of the jet with time, $\psi \equiv \psi \left( t \right)$, which describes the fading nature of this quasi-periodic variability. The combination of these two effects describes the time-dependent trajectory of the blob. As a result, flux variations, as expressed in Eq.~(\ref{eq:flux_expression}), are directly influenced by the evolving dynamics of $\psi\left( t \right)$ and $\phi$, providing a framework for interpreting the observed flux variations over time.

\begin{figure*}
	\centering
        \includegraphics[scale=0.21]{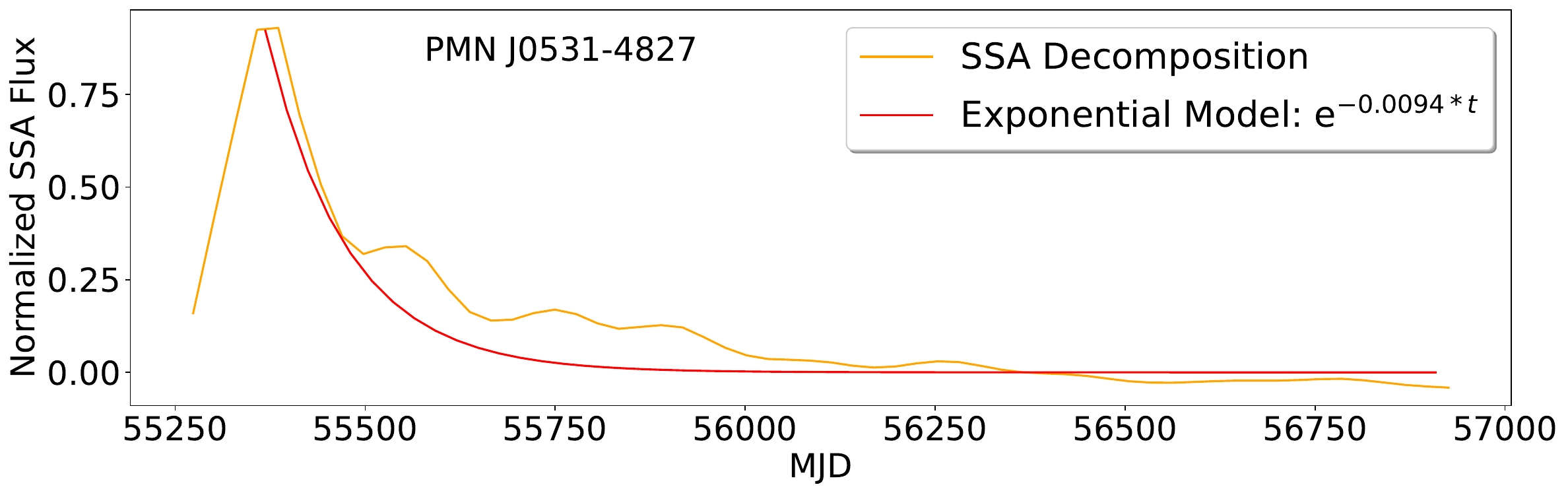}
        \par\vspace{0.5cm}
        \includegraphics[scale=0.21]{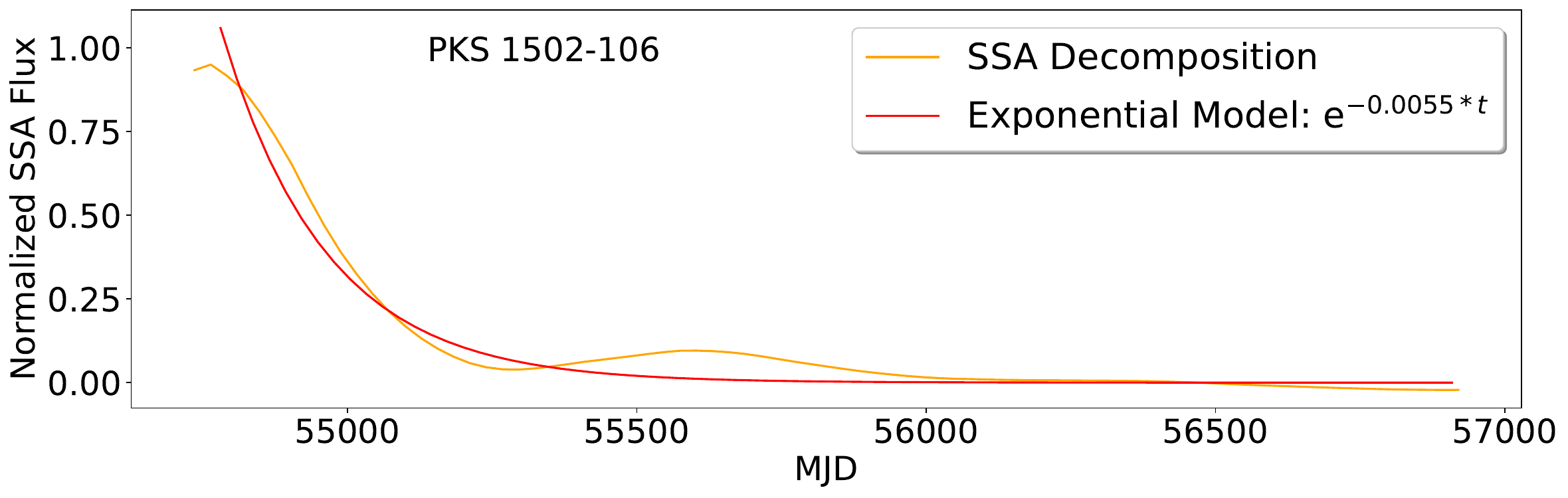}
        \includegraphics[scale=0.21]{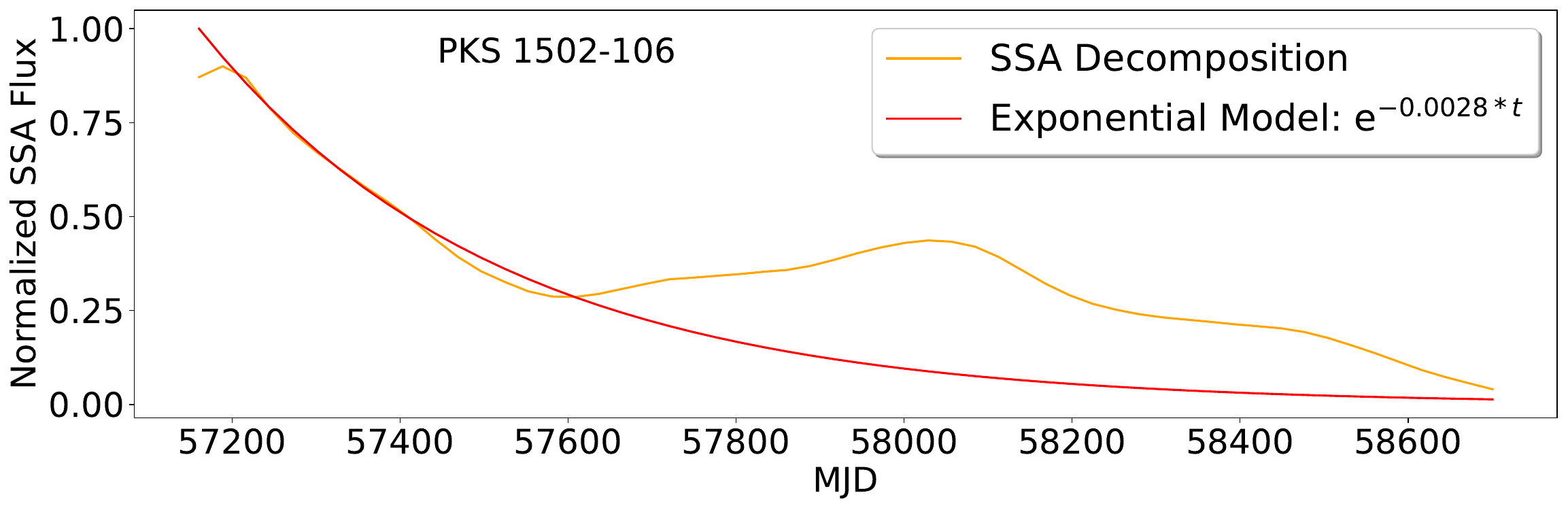}
	\caption{SSA LC decomposition of the studied blazars, revealing the exponential decay component in the signal. Top: PMN J0531$-$4827. Bottom left: First segment of PKS 1502+106. Bottom right: Second segment of PKS 1502+106. The initial value of the exponential index is indicated for each blazar. The SSA decomposing follows the steps detailed in \citet{alba_ssa}.} \label{fig:exponential_ssa}
\end{figure*}

\subsection {Flux Fit}\label{sec:amplitude_fit}

The curved jet model relies on the characteristic exponential decay of oscillations over time as a result of the changing viewing angle as the blob propagates along the jet due to its curved nature. As noted in studies such as \citet[][]{sarkar_curved_jet, anuvab_curve_jet}, in this scenario, QPOs often begin with a flaring state, from which the oscillations exhibit initially high amplitudes that progressively decrease in an exponential manner. This behavior is consistent with the transient nature of QPOs and is visually represented in Fig.~\ref{fig:qpos_lcs}, where the oscillatory amplitudes attenuate over time, consistent with the predictions of the curved jet framework \citep[][]{sarkar_curved_jet, anuvab_curve_jet}.

To evaluate this hypothesis, we express a temporal variation of $\psi$ as $\psi \left( t \right)$=$ae^{-b t}$ following previous studies based on curved jet scenarios \citep[][]{sarkar_curved_jet, anuvab_curve_jet}. This change of the jet viewing angle, coupled with the assumed nature of the motion of the blob, captures the changing amplitude of the QPOs, allowing us to model the observed flux data. Additionally, the model is also dependent on the values of the period of the QPO --- derived from our periodicity analysis --- the Lorentz factor $\Gamma$, the intrinsic spectral index $\alpha$, and the Doppler boosting factor $n$.

To assess the quality of the adopted LC model, we employ the R-squared (R$^{2}$) metric as a measurement of the goodness of the fit. R$^{2}$ can take values between 0 and 1 (or 0 to 100 percent). A higher value is indicative of a better agreement between the model and the real data. We adopt the same criteria as in \citet{hair_r2_2011}, where three categories or agreement levels are defined based on the value of R$^{2}$: ``weak'' for 25\%, ``moderate'' for 50\%, and ``substantial'' for 75\%. These thresholds provide a framework for interpreting the accuracy and effectiveness of the model.

\section {Results and Discussion}\label{sec:flux_analysis}
Here, we test for the two selected sources, the curved jet scenario introduced above, where the helical motion of plasma blobs generates periodic variations in the observed flux, while the changing viewing angle of the jet due to its curvature leads to the exponential decay in the QPO amplitudes. As mentioned in Sect.~\ref{sec:amplitude_fit}, this model also depends on other parameters, in particular $\Gamma$, $\beta$, and $\phi$. 

No specific Lorentz factors in $\gamma$ rays were found for these objects in the literature. Consequently, for the Lorentz factors of PKS 1502+106 and PMN J0531$-$4827, we have assumed a value $\Gamma = 15$ in our model, within typical values often measured for blazars, as reported by \cite[][]{gamma_fsrq_ghisellini}. This can be used to estimate the value of $\beta$ as $\Gamma = 1 / (1 - \beta^{-2})^{1/2}$, resulting in $\beta = 0.99777$. 
The pitch angle, $\phi$, is another variable that is associated with a broad range, typically between $1.5^{\circ}$ and $10.0^{\circ}$ \citep[e.g.,][]{jorstad_pitch_angle, butuzova_pitch_angle}. However, no specific pitch angles were obtained from the literature for these blazars; thus, in alignment with \citet[][]{sarkar_curved_jet, anuvab_curve_jet}, we use a pitch angle of $\phi = 2^{\circ}$.

Moreover, the values of the period $P_{obs}$, the index of the exponential function $b$, the spectral index $\alpha$, and the Doppler boosting index $n$ are left free to optimize the agreement between the model and the data in the fitting process. We modeled the flux LC for each object with Eq.~(\ref{eq:flux_expression}), converging to the values of $P_{obs}$, $b$, $\alpha$ and $n$ that result in the highest value of R$^{2}$. This process is performed systematically with an iterative approach. We consider values of $P_{obs}$ within the uncertainty of the period reported in Sect.~\ref{sec:methodology}. As for the spectral index, we refer to the mean photon index measured by \citet{spectral_index_fsrq_gamma} in a systematic study of Fermi-LAT blazars, corresponding to a spectral index $\sim 1.4$. Therefore, we test $\alpha$ values in the range of 1.2, 1.3, 1.4, 1.5, and 1.6 for the model adopted here. 

For $n$, we consider a range of values based on the types of blazars analyzed in this study. As discussed in Sect.~\ref{sec:sample}, our sample includes an FSRQ and a BL Lac, which exhibit different ranges of values of the beaming factor $n$ due to variations in their dominant $\gamma$-ray emission mechanisms. BL Lacs, characterized by weaker external photon fields compared to FSRQs, predominantly emit through synchrotron radiation and synchrotron self-Compton (SSC) processes \citep[e.g.,][]{ghisellini_2009}. Synchrotron radiation, being more isotropic in the jet's rest frame, exhibits a relatively weak dependence on the Doppler factor. SSC processes introduce a slightly stronger sensitivity to Doppler boosting but are still moderate compared to the external Compton (EC) mechanisms that dominate in FSRQs. Consequently, for BL Lac objects, the Doppler boosting factor is typically $n \approx 2-3$ \citep[][]{dermer_doopler}. FSRQs are characterized by strong external photon fields originating from the broad-line region, accretion disk, and the dusty torus, which typically make EC scattering the dominant emission mechanism at $\gamma$-ray energies \citep[e.g.,][]{ghisellini_2009}. This leads to a higher sensitivity of the observed flux to the Doppler factor, with $n \approx 3-5$ \citep[][]{dermer_doopler}. This heightened sensitivity arises because EC scattering, where seed photons from external sources, such as the accretion disk, broad-line region, or dusty torus, are upscattered by relativistic electrons, enhances energy transformations more significantly than SSC. The external photon field, which typically has a higher energy density than the synchrotron photon field inside the jet, results in stronger Doppler boosting. As a result, the observed flux becomes more sensitive to the Doppler factor, amplifying the effects of relativistic beaming \citep[][]{dermer_doopler}.

\begin{figure}
	\centering
        \includegraphics[scale=0.223]{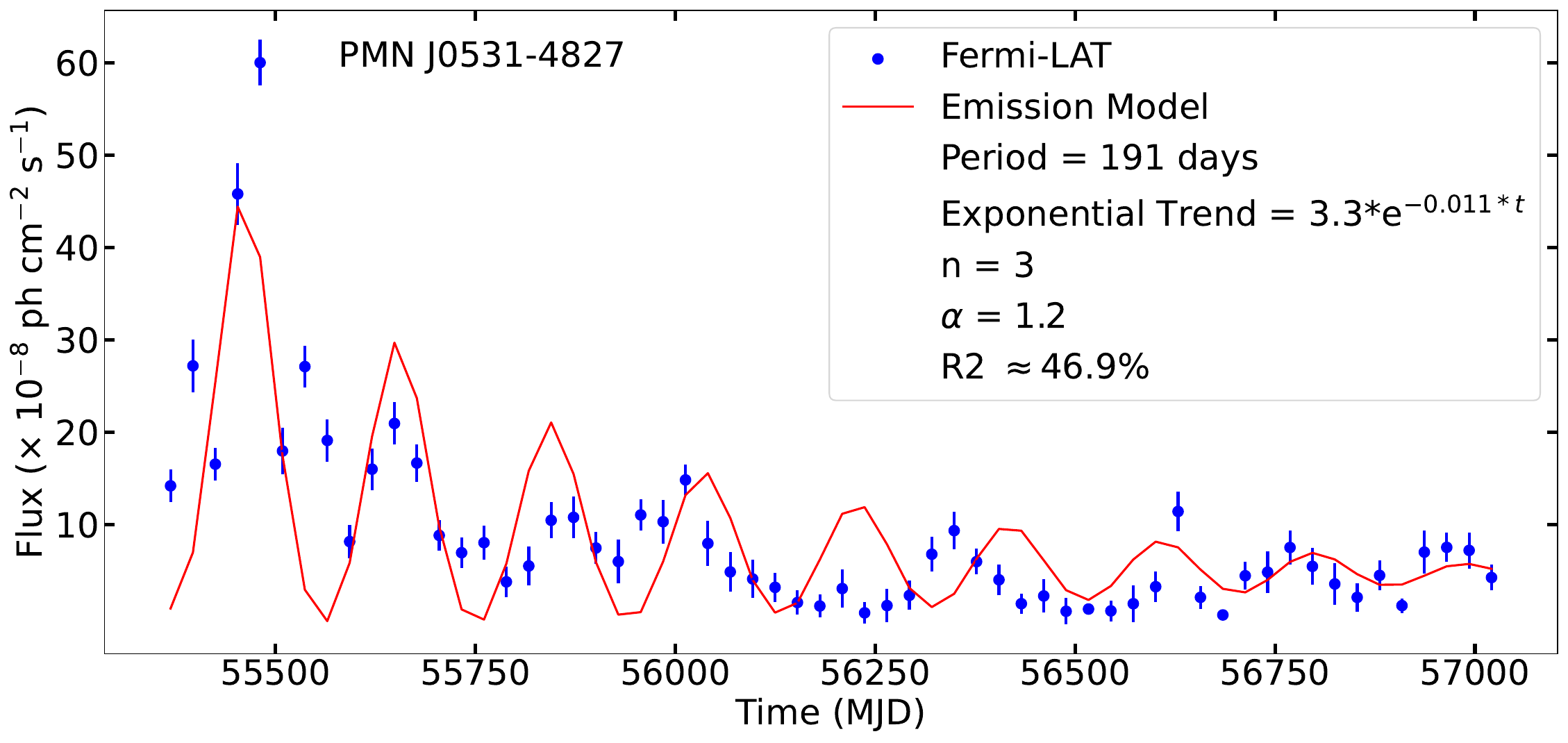}
	\caption{LC from Fig. \ref{fig:qpos_lcs}, illustrating the flux fit of PMN J0531$-$4827. Red lines indicate the emission fit based on Eq.~(\ref{eq:flux_expression}). The plot legends display the fit parameters, showing the derived period and exponential decay estimated to optimize the R${^2}$ value.} \label{fig:flux_models_pmn}
\end{figure}

Finally, the parameters $a$ and $b$ describing the exponential change of $\psi$ are defined by the pair of values that maximize the goodness of the fit R$^{2}$.
The results of the optimization of the model for each blazar are summarized in Table~\ref{tab:fitting_results}, and the resulting LC models are shown in Fig. \ref{fig:flux_models_pmn} and Fig. \ref{fig:flux_models_pks}.

\subsection{PMN J0531$-$4827}
The fit values are shown in Table \ref{tab:fitting_results} and Fig. \ref{fig:flux_models_pmn}. We obtain an R$^2$ of 46.9\%, indicating approximately a ``moderate'' fit based on the classification described in Sect.~\ref{sec:methodology}. The derived values for $n$ and $\alpha$ are 3 and 1.2, respectively. The difference in R$^{2}$ for $n=2$ is 30\% smaller. Regarding the values for $\alpha$, the differences range from 2\% to 8\%. Figure \ref{fig:psi_evolution} illustrates the temporal evolution of the jet's viewing angle, providing key insights into the dynamical changes in its orientation. The inferred period is 191 days. According to the scenarios presented in \citet{rieger_2004} (see Sect.~\ref{sec:model}), this result suggests that the underlying physical mechanism is an intrinsic phenomenon within the jet. These can be associated with helical jet instabilities, plasma turbulence, or internal shock interactions. The helical motion of the jet due to magnetohydrodynamic instabilities can lead to periodic variations in emission \citep[][]{rieger_2004}. At the same time, turbulence in the relativistic plasma may introduce quasi-regular fluctuations in the jet structure \citep[][]{marscher_2014}. Additionally, internal shocks caused by variations in the jet’s velocity could create periodic enhancements in particle acceleration, modulating the observed emission and contributing to the QPO \citep[][]{spada_2001}.  

\begin{figure*}
	\centering
        \includegraphics[scale=0.223]{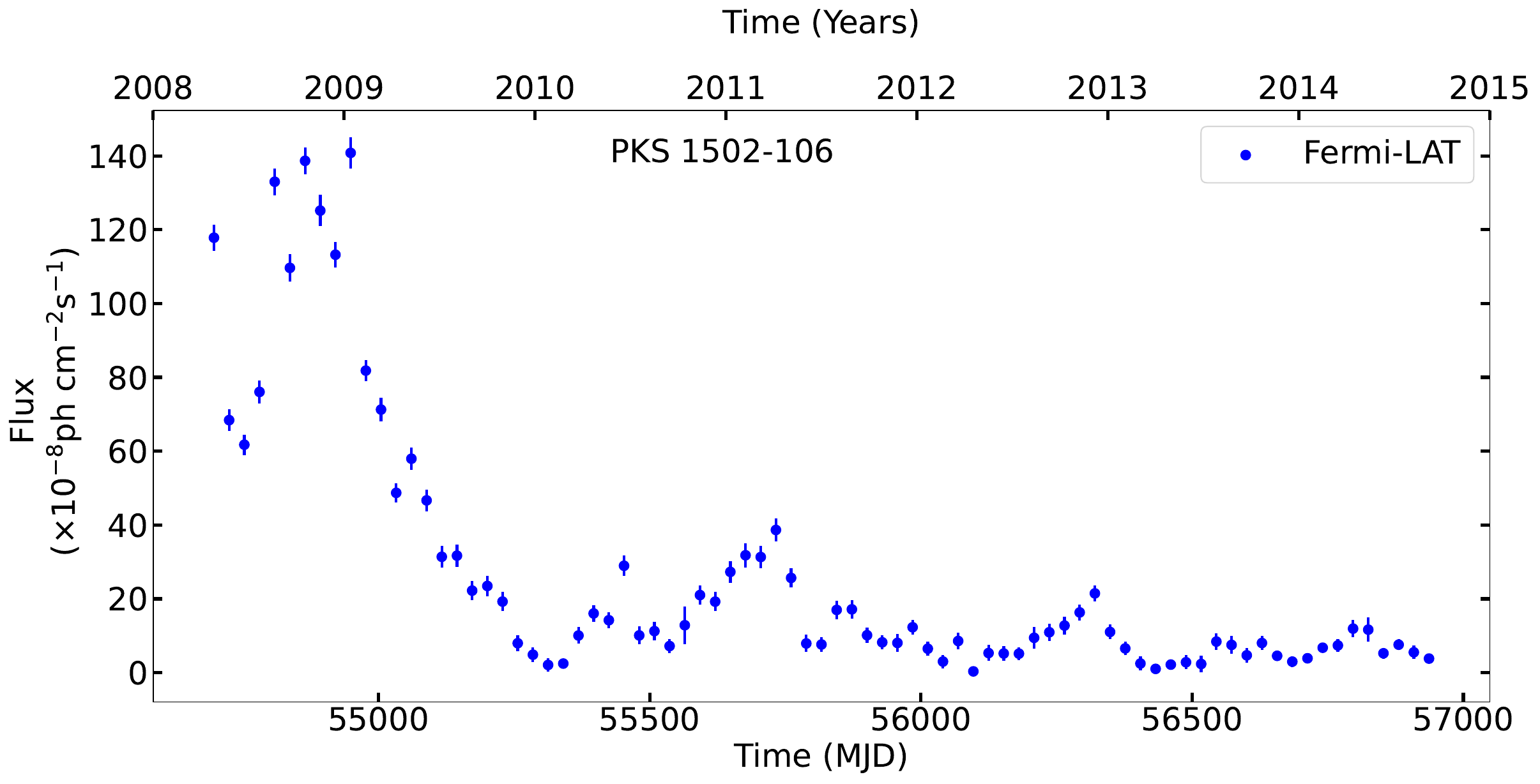}
        \includegraphics[scale=0.223]{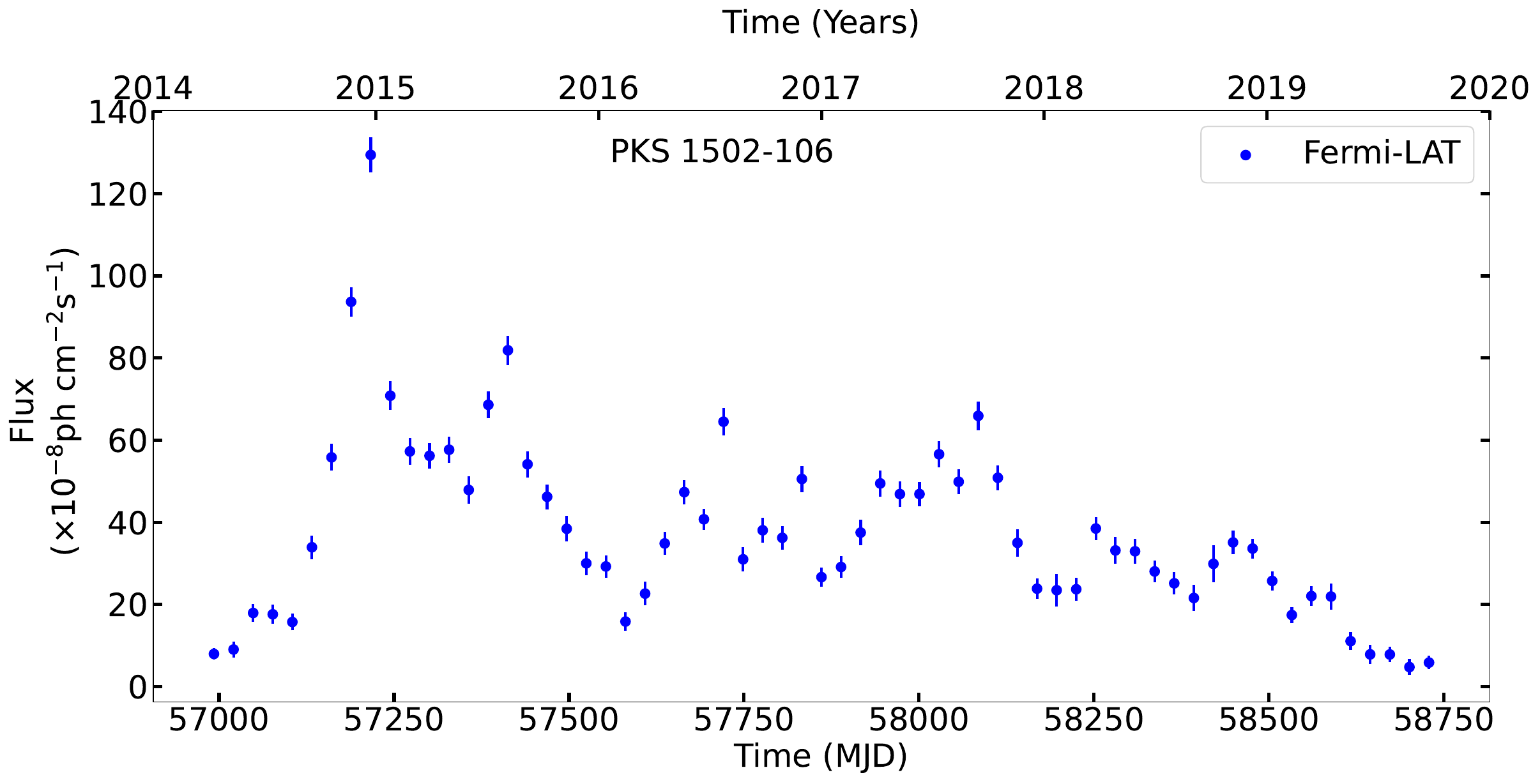}
	\caption{Segments of the LC of PKS 1502+106 analyzed. Left: First segment (54696-56936 MJD). Right: Second segment (56992-58756 MJD).} \label{fig:pks1502_segments}
\end{figure*}
 
\subsection{PKS 1502+106}\label{sec:pks1502_results}
As shown in Fig.~\ref{fig:qpos_lcs}, we identify two distinct segments in the LC that exhibit potentially transient QPOs. The first segment spans MJD 54696–56936, while the second covers MJD 56992–58756 (Fig. \ref{fig:pks1502_segments}). The corresponding oscillatory components extracted via SSA are presented in Fig.~\ref{fig:qpos_ssa}. Each segment is analyzed independently to evaluate its periodic behavior and underlying physical properties. 

\subsubsection {First Segment}
\begin{figure*}
	\centering
        \includegraphics[scale=0.223]{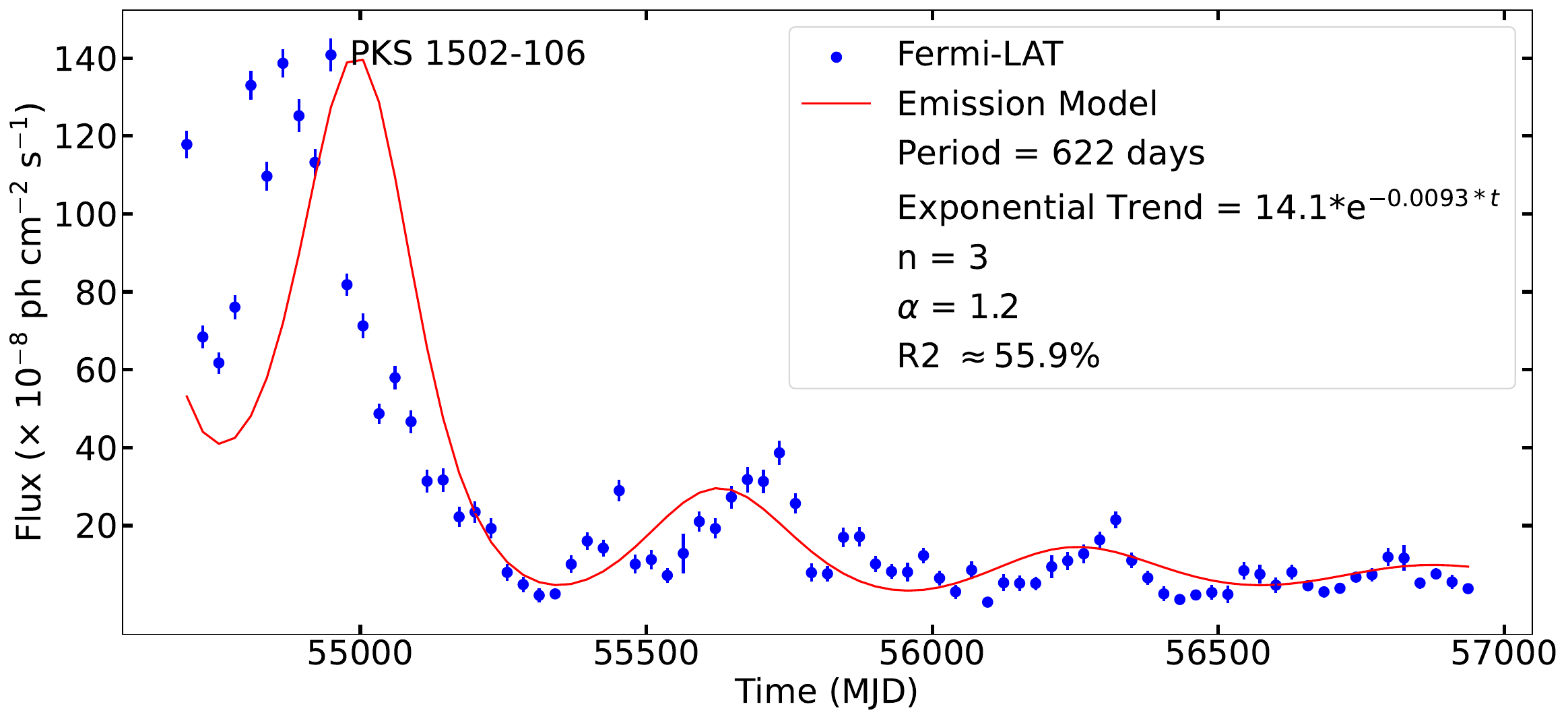}
        \par\vspace{0.5cm}
        \includegraphics[scale=0.223]{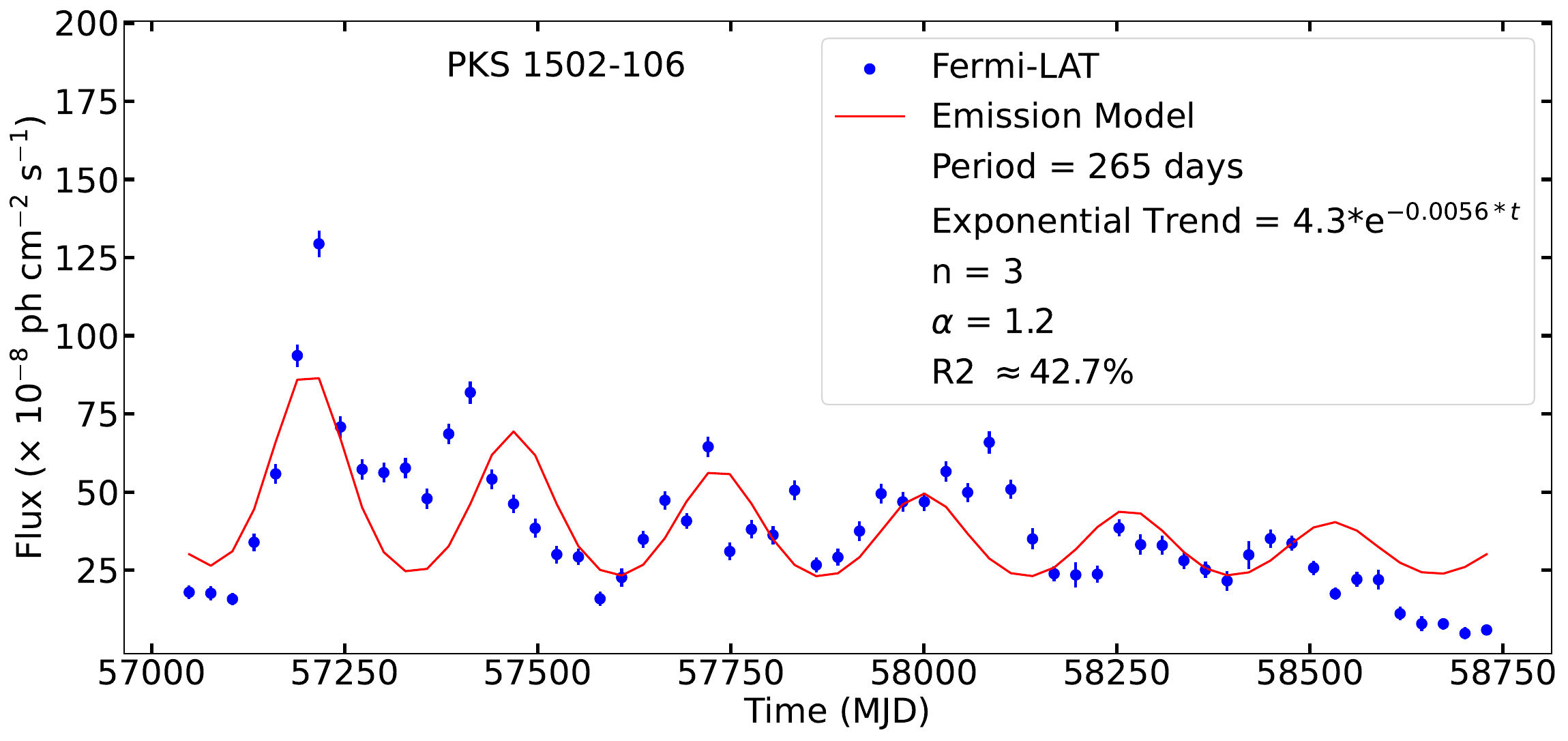}
        \includegraphics[scale=0.223]{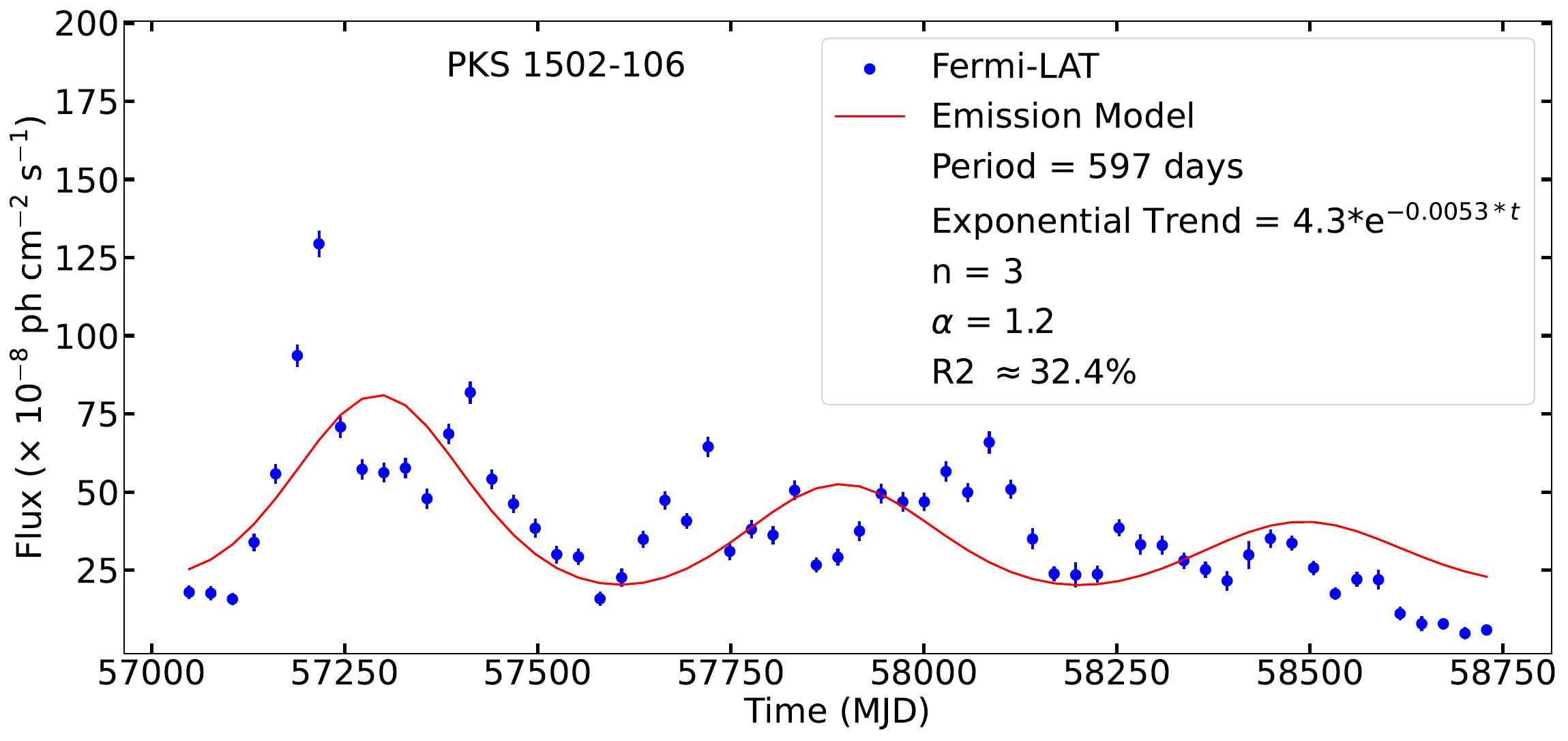}
	\caption{Flux fits of each analyzed segment of PKS 1502+106. Top: First segment (54696-56936 MJD).  Bottom: Second segment (54992-58756). For this segment, two different flux fits are done, according to the two potential periods reported for the QPO. Left: Flux fit considering a period of 283$\pm$47. Right: Flux fit considering a period of $593\pm53$. Red lines indicate the emission fit based on Eq.~(\ref{eq:flux_expression}). The plot legends display the fit parameters, showing the derived period and exponential decay estimated to optimize the R${^2}$ value.} \label{fig:flux_models_pks}
\end{figure*}

The values of the best fit for the first segment of this source are also reported in Table \ref{tab:fitting_results}. Moreover, the derived LC model is shown in Fig. \ref{fig:flux_models_pks}. In this case, the best fit is found with a goodness value R$^{2}=55.9$\%, which can again be considered approximately as a ``moderate'' fit according to the categorization defined in Sect.~\ref{sec:methodology}. The Doppler boosting factor and spectral index converge to values of $n=3$ and $\alpha=1.2$ as for the previous case, with a degradation of R$^{2}$ down to 37\% for larger values of $n$ and 20\% in the case of $\alpha$. Figure \ref{fig:psi_evolution} depicts the temporal evolution of the jet's viewing angle, changing from $\sim0.2^{\circ}$ to $\sim1.4^{\circ}$, highlighting its characteristic exponential decay over time.

\subsubsection {Second Segment}
The second segment of the LC of PKS 1502+106 yields two different candidate periods based on the analysis using SSA and the GLSP. As shown in Table \ref{tab:fitting_results}, both detected periods have relatively low significance ($\leq$1.5). The first period, reported by SSA, is $283\pm47$ days, while the second, at  $593\pm63$ days, is consistent with the period found in the first segment of the LC (see Fig.~\ref{fig:complete_lc_pks_vertical}).

To explore the potential periodic modulation of the flux, we performed two separate exponential fits using each of the candidate periods. The fit associated with the $\approx$300-day period yields an R$^{2}$ value of 42.7\%, while the fit using the $\approx$600-day period gives a lower R$^{2}$ of 32.4\%. The parameters $a$ and $b$ from the exponential model are compatible in both cases, suggesting a similar flux evolution trend regardless of the assumed period. However, based on the higher R$^{2}$ value, the fit associated with the $\approx$300-day period is favored. This suggests that the $\approx$300-day period provides a better description of the observed flux modulation in the second segment of the LC. Figure \ref{fig:psi_evolution} shows the temporal evolution of the jet's viewing angle, from $\sim0.8^{\circ}$ to $\sim1.3^{\circ}$ degrees, significantly different from the viewing angle evolution of the first segment.

\subsubsection {Discussion}
The inferred period of $\approx$600 days in the first segment of the LC may be attributed to the helical motion of a plasma blob along the jet, a scenario commonly associated with binary SMBH systems. As suggested by \citet{rieger_2004}, such long-term periodicities could arise from the orbital motion in a binary SMBH system (see Sect.~\ref{sec:model}). Indeed, the possibility of a binary SMBH at the core of PKS 1502+106 has previously been proposed by \citet{britzen_precission_2016}. In this context, \citet{villata_helical_jet} predicted that the relativistic jet could become bent due to the orbital dynamics of the binary system. Additionally, \citet{raiteri_twested_jet} showed that interactions between the magnetized jet plasma and the surrounding environment can induce a helical twist in the jet structure. This type of geometry may contribute to periodic variability in the observed flux via relativistic Doppler modulation, as also discussed by \citet{liska_precession} in the context of precessing accretion flows.

The second segment of the LC yields two possible periods with relatively low significance, complicating the interpretation. One of these periods ($\approx$600 days) is consistent with the period found in the first segment, providing additional, though tentative, support for the binary SMBH scenario. However, the alternative period of approximately $\approx$300 days results in a better fit to the flux modulation, as indicated by the higher R$^{2}$. Although \citet{rieger_2004} also considered binary systems capable of producing shorter periodicities (1-year-like QPOs), the lack of consistency between the two candidate periods casts doubt on the binary scenario as a unique explanation.

\begin{figure*}
	\centering
        \includegraphics[scale=0.155]{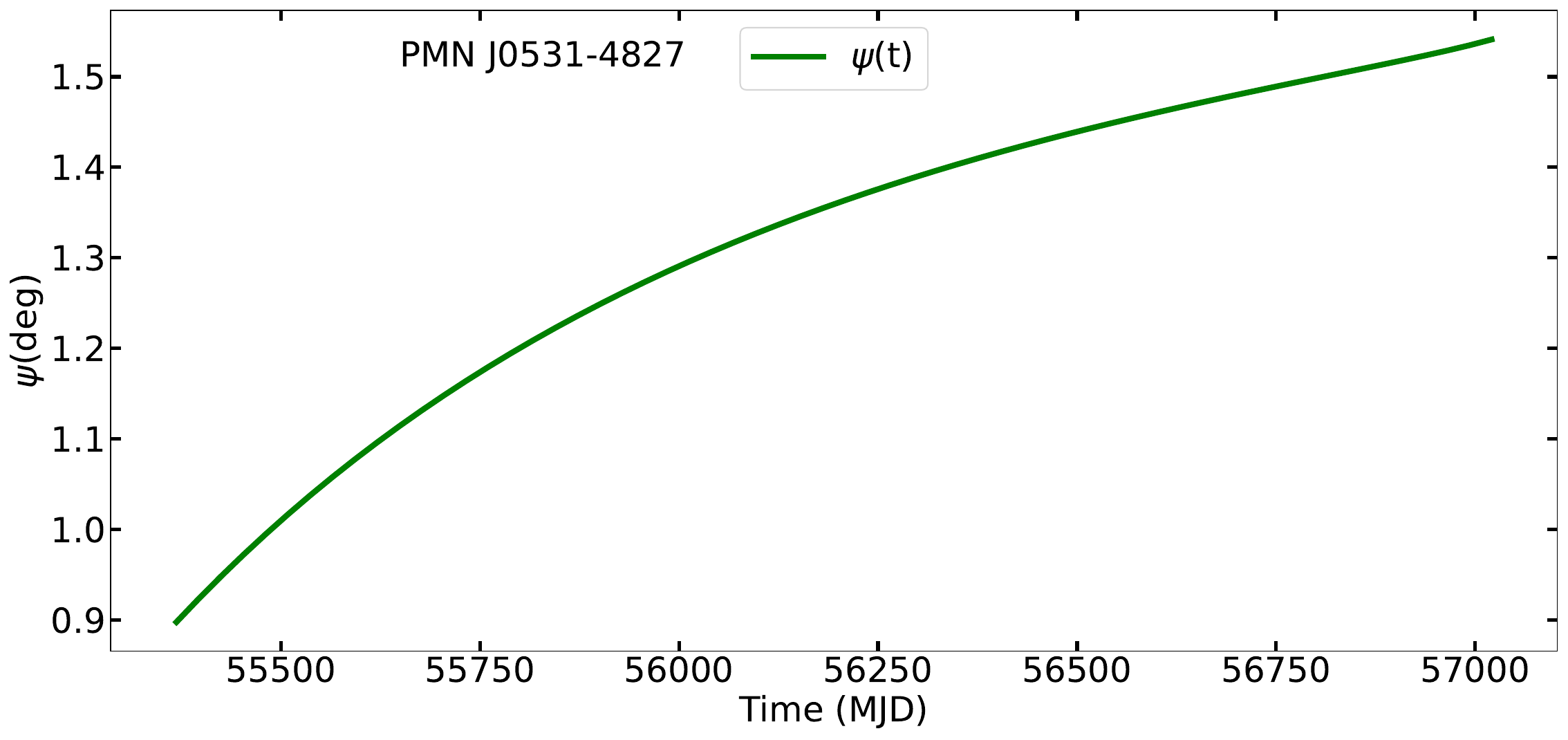}
        \includegraphics[scale=0.155]{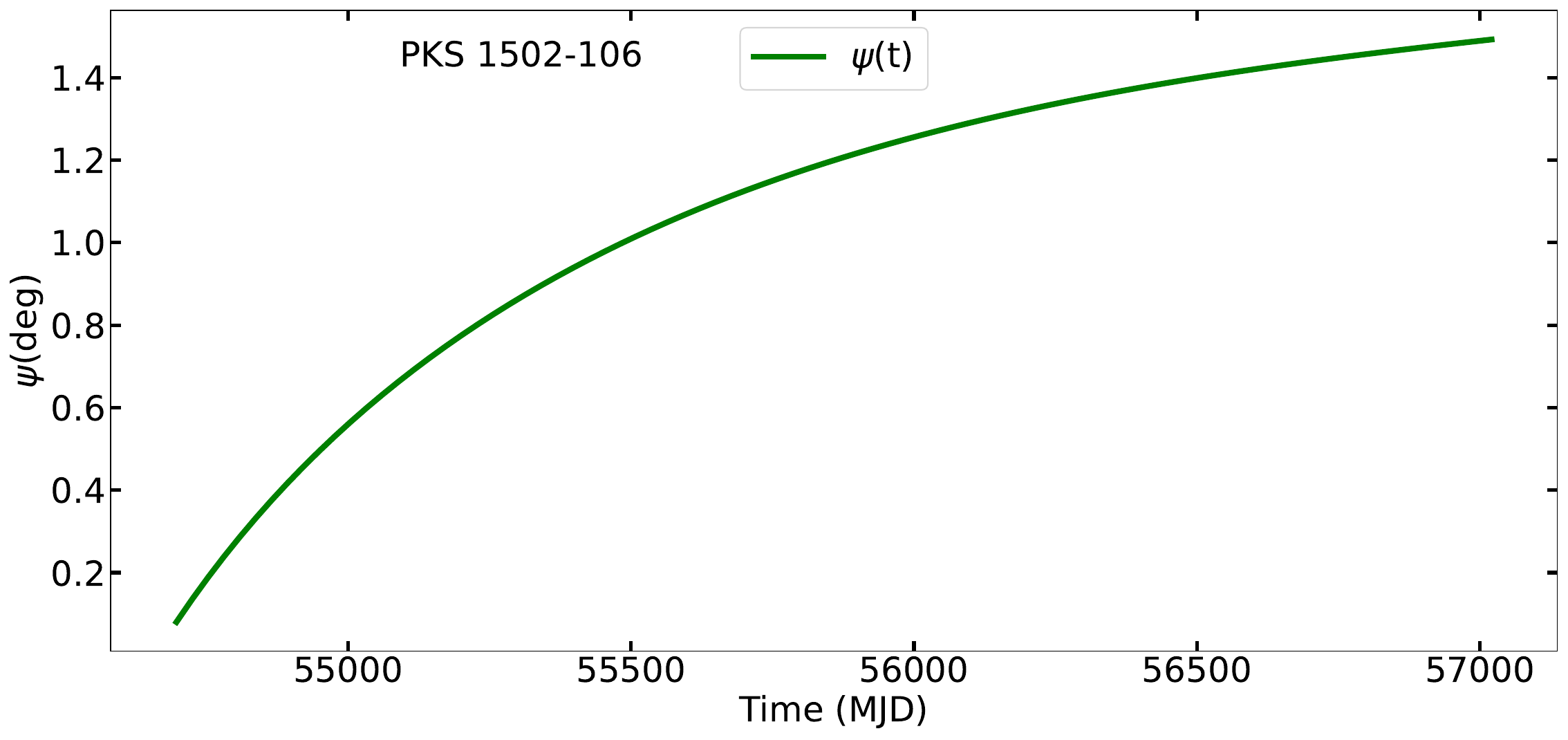}
        \includegraphics[scale=0.155]{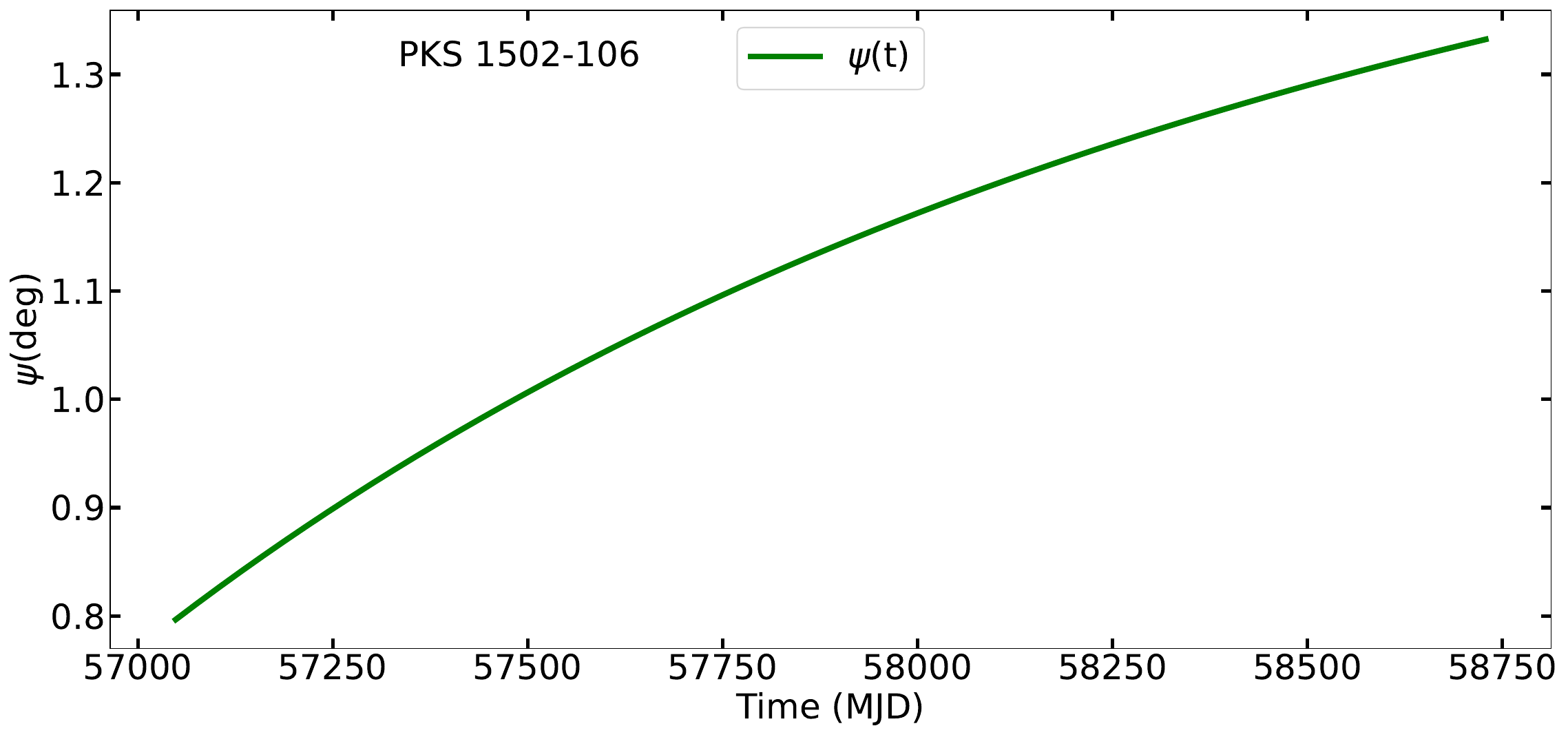}
	\caption{Jet viewing angle ($\psi$) as a function of time. Top: PMN J0531$-$4827.  Center: PKS 1502+106, first segment. Bottom: PKS 1502+106, second segment. We use the parameters with best R$^{2}$ (Table \ref{tab:fitting_results}). The figures for PKS 1502+106 show the differences in the jet viewing angle consequence of the different exponential trend fits (Table \ref{tab:fitting_results}). } \label{fig:psi_evolution}
\end{figure*}

Additionally, the exponential flux trend parameters (specifically, the $a$ and $b$ coefficients) differ between the two segments. This discrepancy suggests a variation in the Doppler boosting factor, which can be interpreted as a change in the jet's viewing angle, as shown in Fig.~\ref{fig:psi_evolution}. 

Despite the inconsistencies found after analyzing both transient QPOs, the binary scenario can not be discarded completely. The binary interaction could result in complex scenarios, resulting in a complex structure in the LC \citep[e.g., ][]{quian_3c279_binary, britzen_oj287_2018}. For instance, in the case of PKS 1502+106, during a different orbital phase, the geometry of the system may cause the jet to sweep past the observer’s line of sight much more rapidly. This could occur if the orbital motion causes a sharper angular curvature in the jet trajectory. In this scenario, the alignment with the observer is brief and less stable. Therefore, the Doppler factor peaks quickly and then declines rapidly as the angle shifts away. The resulting QPO event would appear sharper, with a steeper exponential decay in the LC following the maximum (first transient QPO). In contrast, during another orbital phase, the jet axis may shift more gradually into near-perfect alignment with the observer's line of sight. In this case, these conditions are sustained for leading to a smoother decay in the QPO amplitude after the peak, exhibiting a slow exponential decay in flux (second transient QPO). However, during these phases, other intrinsic jet processes \citep[][]{marscher_2014} may become more prominent in the observed LCs. Such phenomena, like turbulence or localized particle acceleration, can overlap with or obscure the primary QPO signature, complicating the identification of a consistent period between both transient QPOs. \citet{britzen_precission_2016} proposed the presence of a curved jet in PKS 1502+106, whose kinematics appear perturbed by the surrounding, such as narrow-line region clouds or disc-driven winds, further increasing the complexity of the jet.

Therefore, the results for PKS 1502+106 could suggest that a single, curved jet structure may not fully account for the transient nature of the QPOs observed in this blazar. The variability in both period and flux profile points to a more complex scenario, possibly involving multiple physical mechanisms \citep[][]{britzen_precission_2016}, not necessarily considering the binary system. In this context, a possible explanation for the QPOs of this blazar is provided by the interaction between a plasma blob and standing shocks within relativistic jets. As shown by \citet{fichet_increase_trends_shock_2022}, when a high-speed perturbation encounters a recollimation shock, it can trigger a sequence of relaxation shocks that propagate downstream. These relaxation shocks act as secondary dissipation sites, producing delayed emission events or ``flare echoes''. Due to the progressive energy dissipation and the weakening of the shock structure over time, each subsequent emission episode can appear fainter, potentially leading to an exponentially decaying profile. 

An alternative explanation is provided by the model proposed by \citet{wiita_cells_2011}. In this framework, non-axisymmetric magnetic field structures within the jet create localized irregularities. As a shock wave propagates outwards, it interacts with these inhomogeneities, producing fluctuations in the emitted flux. When the shock moves through a helical structure within the jet, the effect is similar to a change in the jet’s direction, resulting in an apparent shift in the viewing angle. If the jet remains approximately cylindrical and the shock velocity is constant, the emission pattern can maintain a consistent period, especially if similar helical structures are encountered sequentially. Furthermore, \citet{wiita_cells_2011} suggests that turbulence behind a propagating shock can also generate quasi-periodic features. In this scenario, a dominant turbulent cell can periodically enhance the emission via Doppler boosting. However, due to the stochastic and decaying nature of turbulence, the amplitude of such QPOs will diminish over time. Changes in the size, strength, or speed of the turbulent cells can also cause variations in the observed period. 

\begin{table*}
\centering
\caption{List of periods obtained by the methodology presented in Sect.~\ref{sec:methodology}, along with the values of various variables that optimized the flux fit based on $R^{2}$.} 
\label{tab:fitting_results}
\begin{tabular}{cccc|ccccc}
\hline
\multirow{2}{*}{Source Name} & Segment & \multicolumn{2}{c}{Period [days]} & \multicolumn{2}{c}{Optimized Fit}  & \multirow{2}{*}{$n$} & \multirow{2}{*}{$\alpha$} & R$^2$ [\%]  \\ 
 &   &  SSA     &  GLSP    &  Period [days]  & Trend fit & & &    \\ \hline
PMN J0531$-$4827 & -- & 188$\pm$39 (2.7$\sigma$) & 213$\pm$46 (1.1$\sigma$) & 191 & \makecell{$a$=$3.3$ \\ $b$=$-1.1\times10^{-2}$} & 3 & 1.2 & 46.9  \\ 
\hline                           
PKS 1502+106 & 54696--56936 & 689$\pm$108 (2.4$\sigma$) & 661$\pm$72 (2.1$\sigma$) & 622 & \makecell{$a$=$13.8$ \\ $b$=$-9.3\times10^{-3}$} & 3 & 1.2 & 55.9 \\  
 & 56992--58756 & 283$\pm$47 (1.2$\sigma$) & 593$\pm$63 (1.6$\sigma$) & 265 & \makecell{$a$=$4.3$ \\ $b$=$-5.6\times10^{-3}$} & 3 & 1.2 & 42.7 \\
 &  &  &  & 597 & \makecell{$a$=$4.3$ \\ $b$=$-5.3\times10^{-3}$} & 3 & 1.2 & 32.4 \\
\hline
\end{tabular}
\end{table*}

\section{Summary} \label{sec:summary}
We have investigated the transient QPOs observed in the $\gamma$-ray emission of blazars PMN J0531$-$4827 and PKS 1502+106 using over a decade of Fermi-LAT observations. Through the analysis of their LCs, we characterize the periods of these transient QPOs, with values of $\approx$200 days for PMN J0531$-$4827. In the case of PKS 1502+106, we analyzed two independent segments exhibiting QPO-like behavior, both suggesting a characteristic period of $\approx$600 days. However, in the second segment, an alternative solution with a period of $\approx$300 days provided a better fit to the flux modulation pattern. 

Concerning the physical origin of these transient QPOs, the oscillation detected in PMN J0531$-$4827 can plausibly be interpreted within the framework of a curved jet scenario.

For PKS 1502+106, we initially explored a curved jet scenario in which the observed quasi-periodic behavior could result from jet precession, potentially induced by a binary SMBH system. However, modeling the two QPO segments requires a discontinuous shift in phase around JD 57000, from $\psi = 1.4^\circ$ to $\psi = 0.8^\circ$, which is difficult to reconcile with the smooth, continuous precession expected in binary-driven scenarios. Moreover, the differing temporal evolution and decay patterns across the two segments are inconsistent with the stable, long-term periodic modulation typically associated with binary-induced jet wobble. These aspects break the continuity of the curved jet model and challenge its physical plausibility in this context. As a more plausible alternative, we propose that the QPOs in PKS 1502+106 arise from internal shocks within the relativistic jet. This interpretation is consistent with recent models of relaxation shocks in perturbed jets and naturally explains the transient, evolving character of the observed QPOs.

\begin{acknowledgements}
P.P. and M.A. acknowledge funding under NASA contract 80NSSC20K1562. J.O.-S. acknowledges founding from the Istituto Nazionale di Fisica Nucleare (INFN) Cap. U.1.01.01.01.009. 

This work was supported by the European Research Council, ERC Starting grant MessMapp, S.B. Principal Investigator, under contract no. 949555, and by the German Science Foundation DFG, research grant “Relativistic Jets in Active Galaxies” (FOR 5195, grant No. 443220636).
\end{acknowledgements}

\bibliographystyle{aa} 
\bibliography{literature.bib}

\clearpage
\begin{appendix}

\section{Figures}
\begin{figure}
	\centering
        \includegraphics[scale=0.223]{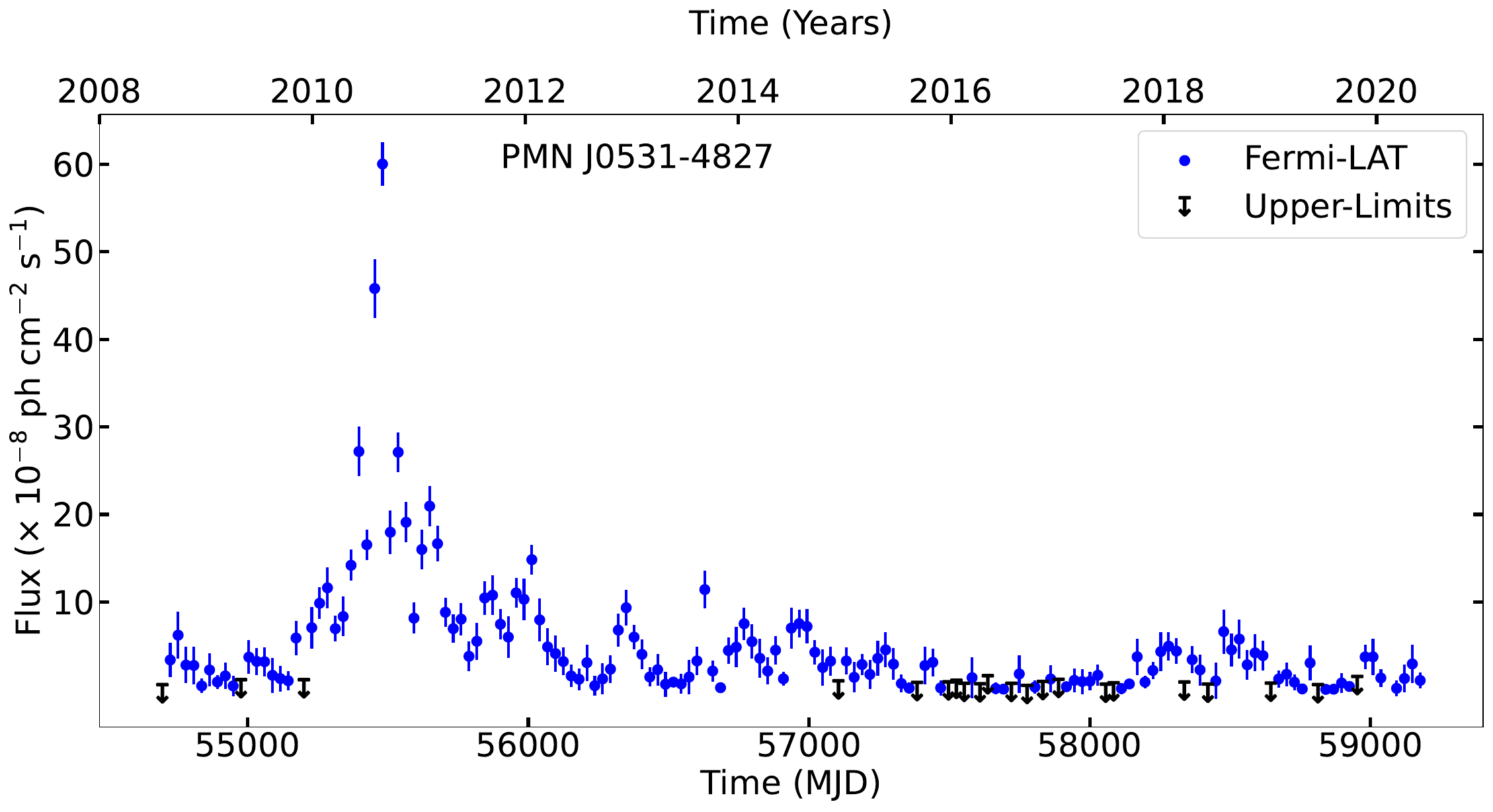}
	\caption{$\gamma$-ray LC of PMN J0531$-$4827.} \label{fig:complete_lc_pmn}
\end{figure}

\begin{figure}
	\centering
        \includegraphics[scale=0.223]{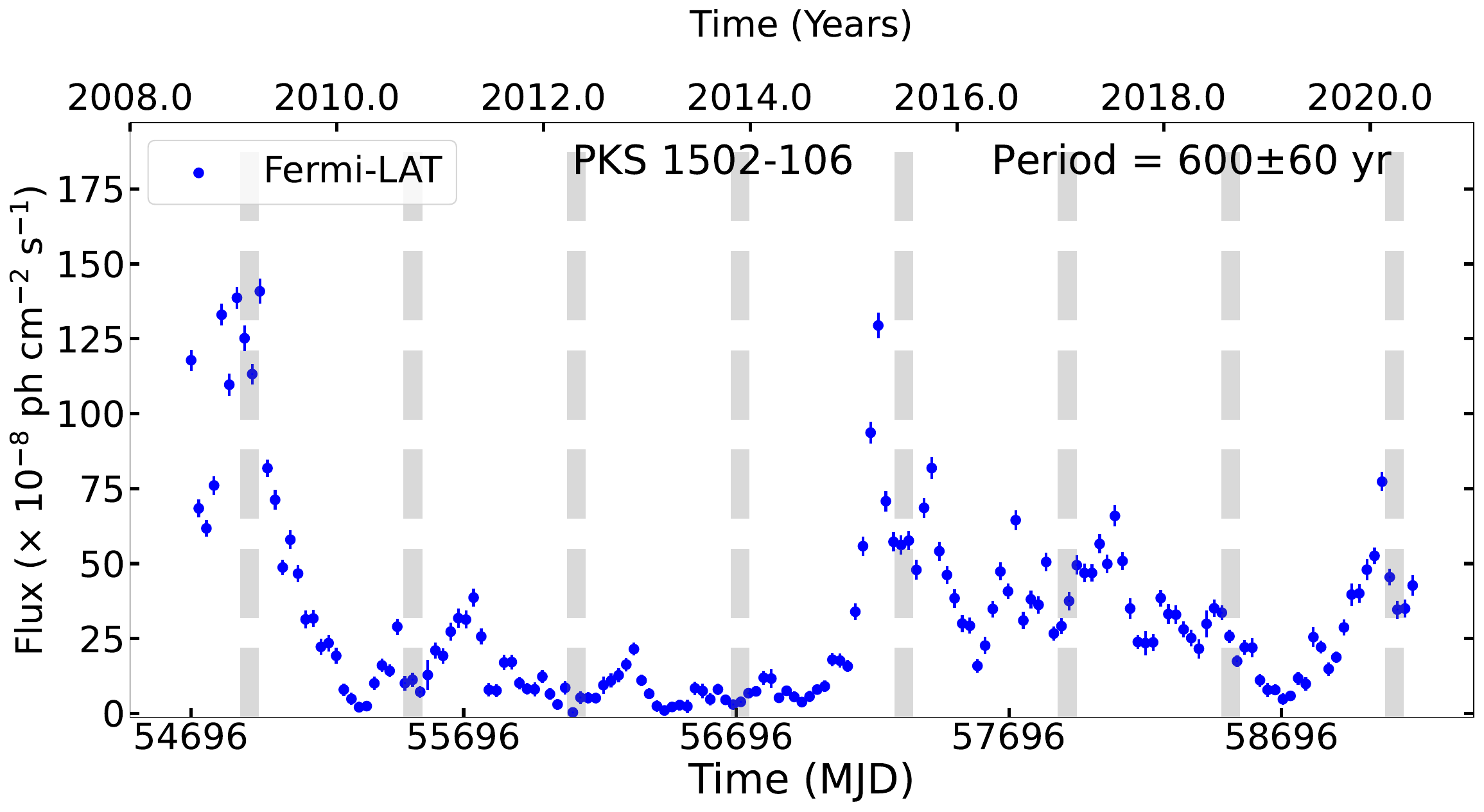}
	\caption{$\gamma$-ray LC of PKS 1502+106. The gray vertical bars are used to approximate the high-flux periods, according to the period $\approx$600 days, which are suggested by our analysis. The width of these gray bars represents the uncertainty associated with the periodic signal, which is an approximation of the results included in Table \ref{tab:fitting_results}.} \label{fig:complete_lc_pks_vertical}
\end{figure}

This section includes the complete $\gamma$-ray LCs of PMN J0531$-$4827 (Fig. \ref{fig:complete_lc_pmn}). Additionally, we include the complete LC of PKS 1502+106, denoting the potential high-flux emission states according to a period of 600 days, an approximation of the results obtained in Sect.~\ref{sec:pks1502_results} (Fig. \ref{fig:complete_lc_pks_vertical}).

\clearpage

\end{appendix}
\end{document}